\PassOptionsToPackage{super,sort&compress}{natbib}
\documentclass{aa}

\usepackage{longtable}
\usepackage{graphicx}  
\usepackage{lipsum}    
\usepackage{lineno}
\usepackage{cuted}
\usepackage{changepage}

\usepackage{amsmath}   
\usepackage{txfonts}  
\usepackage[rgb]{xcolor} 
\usepackage{chngcntr}  
\usepackage{float}
\usepackage{afterpage}
\usepackage{placeins}
\usepackage{makecell}
\usepackage{comment}

\renewcommand{\thetable}{\arabic{table}}

\usepackage[colorlinks=true, linkcolor=blue, citecolor=blue, urlcolor=blue, hypertexnames=false, breaklinks=true]{hyperref}
\usepackage{orcidlink}

\definecolor{myPurple}{rgb}{0.5,0,0.5}

\begin{document} 

   \title{Towards a unified scheme of blazar evolution}

   \author{E. Oukacha
          \and
          Y. Becherini
          }

   \institute{Université Paris Cité, CNRS, Astroparticule et Cosmologie, 75013 Paris, France \\
              \email{enzo.oukacha@gmail.com, yvonne.becherini@apc.in2p3.fr}}

   \date{Received 21 May 2025 / Accepted 2 July 2025}
 
 \abstract
  {Machine learning (ML) and Deep learning (DL) techniques are increasingly being adopted across many fields of astrophysics. With the growing availability of data and refined acquisition methods, these approaches can now be applied to a wide range of tasks, from redshift estimation and light curve variability studies to astrophysical source classification. }
   {In this work, our goal is twofold. Firstly, we wish to classify blazars from the Fermi 4LAC-DR3 catalogue in order to identify the most probable origin of those with currently unknown classifications (BCUs); secondly, we wish to explore the full sample of blazars to investigate the structure and the redshift/luminosity evolution of the blazar population. 
   A particular focus is given to the transition region between Flat Spectrum Radio Quasars (FSRQs) and BL Lacertae objects (BL Lacs), which may provide key insights into the nature and development of the accretion disk activity.
   Building on recent studies, we explore the role of Changing-Look Blazars (CLBs) as potential intermediates in this transition.}
   {We implement a classification approach based on a strong benchmark model (\textit{XGBoost}) and a state-of-the-art foundation model, pre-trained on millions of tabular datasets (\textit{TabPFN}). This constitutes, to the best of our knowledge, the first application of such a pre-trained model to high-energy astrophysics. By extracting the high-dimensional latent space provided by the pre-trained model and by reducing its dimensions, we provide a two-dimensional representation of the blazar population.  
   This enables a nuanced interpretation of the characteristics of sources that lie at the boundary between FSRQs and BL Lacs.}
  {By analysing the reduced latent representation of our data given by the pre-trained model, we identify a clear continuum between FSRQs and BL Lacs, both in terms of high-energy properties and central engine characteristics. This continuous structure reveals a population of sources with intermediate properties, CLBs, which represent a transitional evolutionary stage between FSRQs and BL Lacs. These findings support the scenario of a gradual evolution from FSRQs, with radiatively efficient accretion disks and high Compton dominance, toward BL Lacs characterized by radiatively inefficient flows. }
  {A key outcome of our study is that a single probability score, combined with the pre-trained model’s latent space, can robustly describe any blazar in the sample, offering a new framework for visualizing and interpreting blazar diversity beyond discrete class boundaries. The use of a pre-trained model without the need for domain-specific optimization offers a fast and scalable tool, particularly well-suited for identifying and characterizing ambiguous or transitional sources in current and future blazar catalogues.}

  \keywords{
     BL Lacertae objects: general --
     deep learning -- classification -- foundation model
     methods: statistical -- data-driven analysis
  }

   \maketitle

\section{Introduction}

Blazars are radio-loud (RL) active galactic nuclei (AGN) characterized by a relativistic jet directed along the observer's line of sight. They emit across a broad frequency range, from the radio domain up to very high energies (VHE, \textit{E} > 100 GeV) above 1 TeV. This results in a characteristic double-peaked spectral energy distribution (SED) where the low-to-mid-energy peak is generated by synchrotron emission of relativistic charged particles, while the mid-to-high-energy peak originates from inverse Compton (IC) scattering. Additionally, since Blazars are pointing toward us, they also exhibit high spectral and/or temporal variability. Conventionally, they are further classified into 2 main subtypes: BL Lacertae (BL Lac) and flat-spectrum radio quasars (FSRQ).
The distinction between BL Lac objects and FSRQs is often based on the properties of their optical spectra. One of the main criteria used to differentiate them is the presence (or the absence) of optical emission lines. Objects that are largely featureless in the optical band, showing no strong emission lines or sometimes weak emission and absorption lines, are classified as BL Lacs,  indicative of their jet-dominated nature, \citep{urry_1995_unified}.
As first introduced by \citet{Padovani_1995_BL_LacConnection} and later extended in the Fermi context by \citet{Abdo_2009_SEDFermiBlazars}, BL Lacs are further classified into three subtypes based on the frequency at which their synchrotron emission peaks in their SEDs: low-synchrotron-peaked (LSP), intermediate-synchrotron-peaked (ISP), and high-synchrotron-peaked (HSP) BL Lacs. HSP BL Lacs exhibit synchrotron maxima in the ultraviolet to X-ray regime, indicating highly energetic particle populations \citep{giommi_2015_simplicity}. ISP BL Lacs, representing an intermediate class, have their synchrotron maxima in the optical to ultraviolet range \citep{giommi_2015_simplicity}. LSP BL Lacs, on the other hand, are characterized by synchrotron maxima in the low-frequency regime, from infrared to optical, with inverse Compton attaining its highest intensity in the X-ray to gamma-ray range \citep{giommi_2015_simplicity}.
On the other hand, FSRQs are distinguished by their strong optical emission lines (in contrast to BL Lacs) and pronounced radio emission \citep{urry_1995_unified}. They are typically associated with accretion-dominated systems, where the emission from the central engine is enhanced by thermal contributions from a radiatively efficient accretion disk \mbox{\citep{ghisellini_2011_divide}}. This leads to a more luminous broad-line region (BLR) and a substantial external photon field.
To explain the observed $\gamma$-ray emission in their SED, an external Compton (EC) contribution is generally required. However, for some low- and intermediate-synchrotron peaked (LSP and ISP) BL Lacs, reproducing the observed gamma-ray emission also requires an external photon field, albeit weaker than in FSRQs \citep[e.g.,][]{Bottcher_2013_LeptonicHadronic}. External photon sources for inverse Compton scattering can come directly from the accretion disk or from reprocessed radiation within the BLR and the dusty torus, further enhancing the high-energy component \citep{boettcher_2006_emission}.
As a result of IC, the scattering on the external and internal photon fields, their SED is characterized by a dominant Compton component at high energies, making them some of the most luminous blazars in the $\gamma$-ray domain. On the other hand, it is worth noting that FSRQs tend to be low-peaked blazars, with their synchrotron emission having its maximum at low frequencies, typically around  $10^{13.4} Hz$ \citep{mingaliev_2015_blazars_nupeak}.
In contrast, the absence of strong emission lines in BL Lacs objects suggests that their jets operate in low-accretion-rate environments, where the surrounding photon fields are weaker, reducing the efficiency of external Compton scattering \citep{ghisellini_2011_divide}.

Distinguishing between these classes is essential for understanding both the fundamental physics of jet emission and the characteristics of the environments where these powerful phenomena take place. However, classifying blazars as BL Lacs or FSRQs based on optical and/or multi-wavelength data is not always straightforward. 
A potential evolutionary scenario, initially proposed by \citet{cavaliere_2002_blazar} and supported by \citet{boettcher_2002_blazar_evolution}, suggests that FSRQs and BL Lacs may be connected through a continuous transition rather than being entirely distinct populations. In this scenario, young blazars are found in galaxies with gas- and dust-rich environments, providing abundant material for accretion. 
The abundance of matter in the circumnuclear region sustains a high accretion rate onto the central supermassive black hole, powering a radiatively efficient accretion disk.
The disk emission is characterized by intense thermal radiation, which is subsequently scattered by the BLR, producing strong optical emission lines and a prominent high-energy component dominated by EC scattering. This leads to a high level of Compton dominance, which is typical of FSRQs.
As the accretion rate decreases due to the exhaustion of the circumnuclear material, the system transitions to a radiatively inefficient regime. In this scenario, the ionizing radiation emitted by the accretion disk becomes too weak to sustain the photoionization of gas clouds in the broad-line region (BLR). As a result, the BLR gradually disappears, and the broad optical emission lines, which are produced by recombination in this photo-ionized gas, fade away. Consequently, the contribution from EC scattering diminishes, and although the high-energy component remains Compton-dominated, it becomes comparatively less pronounced. The emission is then primarily driven by synchrotron and synchrotron self-Compton (SSC) processes, as typically observed in most BL Lacs.
This scenario suggests that HSP BL Lacs may have originated from FSRQs, which, after an initial phase of high accretion efficiency, gradually evolved through LSP BL Lacs and ISP BL Lacs before reaching their current state. 
As a result, the transition should appear gradual rather than discrete, highlighting the need for a continuous classification scheme for blazars.
Observational evidence supports the idea that on average, FSRQs tend to be found at higher redshift than BL Lacs \citep{ajello_2022_4lac}, raising the question of whether a cosmological evolutionary link exists between the two populations. Furthermore, recent updates to blazar catalogues \citep{ajello_2022_4lac, massaro_2015_5BZCAT} have identified a growing number of sources classified as blazars of uncertain type (BCUs), highlighting the challenges in defining a clear boundary between these populations.

To address these limitations, more recent approaches rely on machine learning and deep learning techniques to automatically classify blazars based on multi-wavelength data. These methods are usually based on optimized machine learning models or neural networks. The advantage of such models is that they are trained to recognize not only the intrinsic physical properties of FSRQs and BL Lacs but also the relationship between them, allowing the model to subsequently classify blazars of uncertain types.
To apply such approaches effectively, accessing extensive blazar catalogues is essential. 

A typical example is the extensive collection of datasets issued from the Fermi-Large Area Telescope (Fermi-LAT) observations, which has been continuously surveying the entire sky in the GeV energy range (20 MeV - 300 GeV) since its launch in 2008.
By compiling data over the years, the Fermi LAT collaboration has released several generations of high-energy gamma-ray source catalogues, including the 4LAC-DR3: the fourth edition of Fermi's active galactic nuclei catalogue based on the third data release \citep{ajello_2022_4lac}. This catalogue primarily classifies sources as FSRQs, BL Lacs, or BCUs. Additionally, it includes a smaller fraction (less than 2\% of the total sample) of non-blazar AGN sources, such as Radio Galaxies, Seyfert Galaxies, Narrow-Line Seyfert 1 Galaxies, Steep-Spectrum Radio Quasars, Compact Steep Spectrum sources, and generic AGNs. These sources are not included in the present analysis, which focuses exclusively on the blazar population. The detailed statistics of the sources and their types in the 4LAC-DR3 are shown in Table~\ref{tab:source_stats}.
\begin{table}[t]
  \centering
  \caption{Summary of source classifications in the 4LAC-DR3 catalogue.}
  \label{tab:source_stats}

  \begin{tabular}{lc}
    \hline
    \hline
    Type & Count \\
    \hline
    FSRQ & 792 \\
    BL Lac & 1458 \\
    BCU & 1493 \\
    non-blazar AGN & 71 \\
    TOTAL & 3814 \\
    \hline
  \end{tabular}

  \tablefoot{For each class, we group upper-case (identified) and lower-case (associated) labels, as our classification relies solely on physical and spectral properties, independent of positional information.}
\end{table}

As a side note, it is worth noting that in Fermi's blazar catalogues, the source classification appears in either upper- or lower-case letters. Lower-case letters denote associations, typically based on spatial coincidence with known counterparts, while upper-case letters indicate identifications supported by additional evidence, such as correlated variability across multiple wavelengths. In our study, since the classification is based purely on physical and spectral properties, and not on positional criteria, we treat associated and identified sources equally. This allows us to group them within each class (e.g., fsrq and FSRQ) as shown in Table~\ref{tab:source_stats}, without introducing bias linked to the identification confidence level.

Following this classification, several recent studies tried to categorize blazars of uncertain type as either FSRQs or BL Lacs. For instance, \citet{agarwal_2023_classif_blazars} used a voting ensemble strategy combining multiple classification algorithms to classify 943 BCUs from the 4LAC-DR3 into 610 BL Lacs and 333 FSRQs. Similarly, \citet{bhatta_2024_classif_blazars} used supervised and self-supervised learning approaches with neural networks and specific weight initialization strategies to classify 1115 BCUs from the same catalogue, assigning 820 to BL Lacs and 295 to FSRQs with their best model.
These recent studies primarily focused on optimizing the classification performance of BCUs using various models (or model ensembles) and learning strategies.

In contrast, in this study, our goal goes beyond a simple two-class separation. Rather than simply assigning a label to each BCU, we aim to gain a deeper understanding of the global structure of the blazar population, and in particular, to investigate the continuum between FSRQs and BL Lacs.
By analysing the internal feature latent space of our best-performing model, we seek to determine whether a continuous transition exists between these two classes, and whether some sources, including BCUs, occupy a stable intermediate region. 
If so, this would suggest that a subset of sources are misclassified because they represent genuine transitional objects with intermediate physical properties.

In this context, several recent studies have explored the possibility of such transitions, especially for sources whose properties deviate from those of typical FSRQs or BL Lacs.

Recent literature often refers to such transitional behaviour as changing-Look AGN, which can describe two distinct phenomena: 
\begin{enumerate}
    \item Changing-obscuration AGN, where variations in the column density along the line of sight (e.g., due to moving clouds or outflows) cause the broad emission lines to appear or disappear;
    \item Changing-state AGN, where intrinsic changes in the accretion rate lead to the emergence or disappearance of the continuum emission and broad lines \citep{Ricci_2023_ChangingLookAGN}.
\end{enumerate}

A well-documented example of changing-Look AGN is the blazar OQ~334, which underwent rapid transitions between FSRQ-like and BL Lac-like states between 2018 and 2020 \citep{Ren_2024_OQ334}. During this period, significant variability was observed in its Fermi light curves and SED, including: the equivalent width of broad emission lines (above 5~\AA{} during FSRQ-like phases), the energy density of the external photon field ($U_{\mathrm{EXT}}$), the luminosity of the accretion disk ($L_{\mathrm{d}}$), and the gamma-ray photon index ($\Gamma$), which was harder ($\Gamma < 2$) during FSRQ phases. These rapid changes were interpreted as shifts between a standard Shakura-Sunyaev accretion disk (SSD) \citep{Shakura_Sunyaev_1973} and an advection-dominated accretion flow (ADAF) regime \citep{yuan_2014_hot_accretion, prandini_2022_blazar_sequence}. However, the physical mechanisms enabling such rapid accretion mode changes, on timescales of days to weeks, remain uncertain.

In parallel, \citet{kang_2023_clb} introduced the concept of changing-look blazars (CLBs) to describe sources that undergo a much slower transition between FSRQ and BL Lac states. In their follow-up study, \citet{kang_2024_clb} analysed eight observational features (among which there are the redshift, the photon index, and the disk luminosity) and found that the density distribution of CLBs in feature space lies between that of FSRQs and BL Lacs. Interestingly, these CLBs exhibit Eddington-scaled accretion disk luminosities in the range $\lambda \in [-1.66, -1.59]$, consistent with the critical threshold expected for a transition from radiatively efficient to inefficient accretion, and supporting the evolutionary scenario proposed by \citet{cavaliere_2002_blazar} and \citet{boettcher_2002_blazar_evolution}. Thus, the CLB definition adopted in \citet{kang_2023_clb, kang_2024_clb} follows a long-term evolutionary view of blazar development, in contrast with rapid state-switching sources such as OQ~334.

This long-term evolutionary scenario serves as the foundation of our analysis, especially because we focus on the 4LAC-DR3 catalogue, which encompasses 12 years of Fermi-LAT data. In contrast, investigating rapid transitions in the accretion mode, such as the case proposed for OQ 334, would require a dedicated time-resolved spectral analysis on a period-by-period basis, which falls beyond the scope of the present work.

To effectively capture such intermediate or evolving sources, we argue that flexible and generalizable models are needed. In particular, foundation models trained on large-scale datasets offer an efficient way to transfer knowledge and adapt to diverse classification or regression tasks across observational catalogues.
In this study, we adopt a Transfer Learning approach to classify blazars in the 4LAC-DR3 catalogue. Crucially, our framework outputs probabilistic predictions and uncertainty estimates, rather than hard labels. 
In summary:
\begin{enumerate} 
    \item We optimize our models to learn the physical differences between FSRQs and BL Lacs using well-identified sources. \item We subsequently deploy the classifiers on the BCUs, assigning each object a probability of belonging to the FSRQ or BL Lac class.
\end{enumerate}
This approach not only refines the classification of BCUs, but also provides new insights into the structure of the blazar population. By visualizing the latent space of our model, we explore potential transitional regions and investigate whether BCUs, and possibly CLBs, occupy a well-defined intermediate region in parameter space.
Combining class probabilities with latent space representation demonstrates that our framework enables a more nuanced and physically meaningful classification of blazars. This allows us to go beyond rigid SED-based schemes and move toward a continuous, data-driven understanding of blazar diversity. 

The results of our classification, along with the two-dimensional representation of the learned latent space, are compiled into an online catalogue that is publicly available on zenodo\footnote{\hyperlink{DOI}{10.5281/zenodo.15577932}}. This resource is intended to support further studies and facilitate reproducibility.

Our work is structured as follows: We begin with a description of the catalogue in Section \ref{4LAC-DR3}. The classification pipeline for BL Lacs and FSRQs and the application of the model to BCUs are presented in Section \ref{classif_pipeline}. In Section \ref{results}, we present our results in the context of a blazar evolutionary scenario. Finally, our conclusions are given in Section \ref{discussion}.

\section{The fourth catalogue of active galactic nuclei detected by the Fermi-LAT (4LAC-DR3)} \label{4LAC-DR3}

The dataset of blazars we consider in this work was obtained from the 4LAC-DR3 catalogue \citep{ajello_2022_4lac}, published by the Fermi-LAT Collaboration, which includes both high-latitude (\(b > 10^\circ\)) and low-latitude sources (\(b < -10^\circ\)). 
The 4LAC dataset is derived from the third data release of the 4FGL \citep{abdollahi_2022_4fgl} catalogue, based on 12 years of gamma-ray data observations by the Fermi satellite ($E_\gamma > 50\;$MeV), and comprises 3814 AGN, the majority of which are blazar-like.
As mentioned in the introduction, blazar-like sources in the 4LAC dataset are classified as either FSRQs, BL Lacs, or BCUs.

The first step of our study is to optimize machine-learning models capable of distinguishing an FSRQ-like blazar from a BL Lac, and then to classify BCUs by deploying them.
To this end, we examine all the available features for the FSRQ and BL Lac samples carrying physical properties of the SED and the information about their variability.

These features include the synchrotron peak frequency (\texttt{nu\_syn}) and its associated flux density (\texttt{nuFnu\_syn}); the energy of the inverse‑Compton peak (\texttt{HE\_EPeak}); the pivot energy (\texttt{Pivot\_Energy}); the $\gamma$‑ray spectral indices obtained from different spectral shapes; and the integral photon flux from 1 to 100\,GeV (\texttt{Flux1000}).
The spectral index is derived from two main models: \texttt{PowerLaw}, and \texttt{LogParabola} obtained by fitting the SED in the $Fermi$ range with a powerlaw and log-parabola model, respectively.
The 4LAC-DR3 catalogue provides, for each source, the best-fit model of its SED, selected among \texttt{PowerLaw}, \texttt{LogParabola}, and \texttt{PLSuperExpCutoff4} models. However, within our final sample, only 6 sources are best described by the \texttt{PLSuperExpCutoff4} model. Due to this under-representation, we chose not to include features specific to this model in our analysis.
Given the strong variability of blazars, flux-related parameters provide valuable information to refine the classification. The 4LAC catalogue provides two temporal‑variability diagnostics: (i) the significance metric $\texttt{Variability\_Index}$, and (ii) the fractional variability parameter $\texttt{Frac\_Variability}$, computed from the excess variance after accounting for both statistical and systematic uncertainties. In the 4LAC-DR3 a source is considered variable at 99\% confidence if the \texttt{Variability\_Index} is larger than a value of $21.67$. 

\begin{table*}[t!]
\centering 
\caption{Key 4LAC-DR3 features for our classification analysis}
\label{tab:main_features} 
\renewcommand{\arraystretch}{1.4}
\centering
\begin{tabular}{p{2.85cm} p{1.9cm} p{9cm} p{1.5cm}}
\hline
\hline
\textbf{Feature} & \textbf{Unit} & \textbf{Description} & \textbf{Coverage} \\
\hline
\texttt{nu\_{syn}} & Hz & Synchrotron-peak frequency (observer frame) & 74\% \\
\texttt{nuFnu\_syn} & erg cm$^{-2}$ s$^{-1}$ & SED at synchrotron peak frequency & 74\% \\
\texttt{HE\_EPeak} & MeV & Energy of high-energy SED peak (observer frame) & 86\% \\
\texttt{HE\_{nuFnuPeak}} & erg cm$^{-2}$ s$^{-1}$ & SED at the high-energy-peak frequency & 86\% \\
\texttt{Highest\_energy} & GeV & Energy of the highest-energy \texttt{ULTRACLEANVETO} photon with association probability $P > 0.95$ & 100\% \\
\texttt{Energy\_Flux100} & erg cm$^{-2}$ s$^{-1}$ & Energy flux in the 100 MeV--100 GeV range from spectral fitting & 100\% \\
\texttt{Flux1000} & cm$^{-2}$ s$^{-1}$ & Photon flux from 1 to 100 GeV & 100\% \\
\texttt{Pivot\_Energy} & MeV & Pivot energy of spectral fit & 100\% \\
\texttt{PL\_{Index}} & – & Photon index when fitting with PowerLaw & 100\% \\
\texttt{LP\_{Index}} & – & Photon index at \texttt{Pivot\_Energy} when fitting with LogParabola & 100\% \\
\texttt{Frac\_Variability} & – & Fractional variability (excess variance) & 74\% \\
\texttt{Variability\_Index} & – & Variability index from Fermi-LAT light curves & 100\% \\
\hline
\end{tabular}
\end{table*}
These features are summarized in Table~\ref{tab:main_features} along with a short description of their meaning.
As seen in this table, some of these features are only available for a fraction of the total number of sources. This is indicated in the coverage column, which shows, for instance, that \texttt{nu\_syn}, \texttt{HE\_EPeak}, and \texttt{Frac\_Variability} are available for only 74\%, 85\%, and 74\% of the sources, respectively.
Since the purpose of this study is to maximize the number of classified sources, there are two possible approaches : 
\begin{itemize}
    \item Restricting the analysis to features available for most sources, sacrificing potentially informative variables.
    \item Instead of discarding features, use a model that natively handles missing data without requiring imputation, thereby preserving valuable information from under-represented features.
\end{itemize}
Given the constraints of previous approaches that relied on datasets with no missing values \citep{agarwal_2023_classif_blazars}, we explore the second strategy, using models capable of directly managing missing values.
Additionally, when working with Fermi data, it is crucial to account for potential issues in the analysis to avoid introducing analysis-related biases. These are indicated through binary analysis flags, which signal specific problems in the data. For instance, a flag can be set if the flux ($>1\; \mathrm{GeV}$) varies significantly (by more than $3\sigma$) when changing the diffuse model. Other flags include poor localization quality or issues with the SED fit \citep[see][Section 3.7.3]{abdollahi_2020_4fgl}. To ensure a clean and unbiased blazar sample for our analysis, we include only sources that have not triggered any analysis flags for the model optimisation. In the 4LAC-DR3 catalogue, this corresponds to sources with a flag value of 0, reducing the sample to 3120 blazar-like sources.
Finally, before optimising machine learning models, examining the distributions of the selected features and their relationships is crucial to ensure they provide meaningful information regarding the class separation. 
The feature space of the main features of the 4LAC-DR3 is shown in Fig~\ref{fig:pairplot_4LAC_features} of Appendix \ref{feat_visualization}.
As expected, a clear distinction between FSRQs and BL Lacs is already visible, particularly concerning \texttt{nu\_syn}, \texttt{HE\_EPeak}, \texttt{PL\_Index}, \texttt{Pivot\_Energy}, and \texttt{Frac\_Variability}.
On the other hand, BCUs appear to form a transitional class between FSRQs and BL Lacs, as their distribution in the feature space overlaps with both populations. 
Further details about data preparation are given in section \ref{FeatureOptimization}.

On a related note, \citet{ajello_2022_4lac} shows a clear distinction in redshift distributions between the two classes: FSRQs have a redshift (\texttt{z}) distribution extending well beyond $\texttt{z} = 2$ with a peak above $\texttt{z} = 1$. On the other hand, BL Lacs are mostly found at lower redshift, with a distribution peaking around $\texttt{z} \approx 0.3-0.4$. However, this observed distribution may be partially influenced by selection effects, as the lack of strong optical emission lines makes it difficult to determine the spectroscopic redshift of many BL Lacs, and fainter high-redshift BL Lacs could remain undetected. Additionally, only a small fraction of sources in the 4LAC-DR3 catalogue have a measured redshift ($\approx 51\%$), which limits its usability as a classification feature. Consequently, we chose not to incorporate the redshift into the classification process. Nonetheless, when available, this information will be valuable to further interpret the final results.

Another valuable resource for understanding the link between different blazar types is the catalogue of central engine properties for Fermi blazars compiled by \citet{paliya_2021_central_engines}. This dataset provides key quantities such as the black hole mass ($M_{\mathrm{BH}}$), accretion disk luminosity (\texttt{Ld}) and Compton dominance (\texttt{CD}) defined as the ratio of inverse Compton to synchrotron peak luminosities. These parameters are especially relevant in the context of an accretion-mode transition between FSRQs and BL Lacs. However, this catalogue only covers about 27\% of the sources in the 4LAC-DR3. To avoid bias and missing data propagation, we decided not to include these parameters in the classification process either. 
However, as for the redshift, this information about the central engine will be valuable when it comes to the interpretation of the final results.

\section{Designing the blazar classification pipeline}\label{classif_pipeline}

\subsection{Supervised learning for blazar type prediction}
\label{classif_models}

When applying machine learning (ML) and deep learning (DL) techniques, it is considered good practice to begin with a simple algorithm and only move to more complex models if necessary. Using such simple models, often referred to as baseline models, provides a useful benchmark to compare the results of more sophisticated methods. 
We will then proceed in the evaluation of the classification performance through a more advanced approach offered by pre-trained foundation models.
In the scope of our study, we will be training models in a supervised manner, i.e, using a set of features $x_{i}$ derived from the catalogue described in section \ref{4LAC-DR3}, to predict a target variable $y_{i}$, which corresponds to the class of blazar (either FSRQ or BL Lac).

\subsubsection{A baseline model using extreme gradient boosting (\textit{XGBoost})}

As a baseline model for our classification task, we use \textit{XGBoost} \citep{Chen_2016_xgb}, an algorithm based on decision tree ensembles and boosting strategies. 
A decision tree is a hierarchical structure that splits data based on discriminative features. 
Each node in the tree represents a condition on a feature, while the leaves contain the final predictions. 
However, using a single tree has several drawbacks, including over-fitting, as deep trees fit too closely to the training data, preventing the model from generalizing well on unseen data.

To address these issues and improve the performance, some models combine multiple trees. This ensemble of trees can be either trained independently on different data subsets, with their predictions being averaged, with a procedure called \textit{Bagging}, used for instance in the \textit{Random Forest} algorithm \citep{breiman_2001_random}, or they can be trained sequentially, with each new model correcting the errors of the previous one, with a procedure called \textit{Boosting}. This latter approach is used in \textit{XGBoost} and often leads to higher accuracy than bagging, even with fewer trees.

The primary objective of our classification task is to minimize the log-loss function, which quantifies the difference between predicted probabilities and actual class labels.

\textit{XGBoost} stands out due to optimizations such as depth-wise tree growth, where trees expand vertically first, which allows it to efficiently capture complex feature interactions, often outperforming \textit{CatBoost} \citep{prokhorenkova_2018_catboost}, \textit{Random Forests} \citep{breiman_2001_random}, \textit{Support Vector Machines} \citep{cortes_1995_support} and \textit{Multi-Layer Perceptrons} \citep{hornik_1989_multilayer} on structured data.

It also incorporates advanced regularization techniques to prevent over-fitting and is optimized for CPU vectorization, GPU acceleration, and multi-threading, making it both fast and robust. 

A key advantage of \textit{XGBoost} for our particular case is its native handling of missing values.
Unlike many other machine learning models that require explicit imputation, \textit{XGBoost} learns the optimal default split direction for missing values during training. Specifically, when a feature is missing, the model determines whether the instance should be sent to the left or right leaf in a way that maximizes information gain. This mechanism ensures that missing values are efficiently accounted for within the decision trees.

While previous studies \citep{agarwal_2023_classif_blazars, bhatta_2024_classif_blazars} have explored the classification of FSRQs and BL Lacs using custom neural network architectures or ensembles of classifiers, we adopt \textit{XGBoost} as a strong and well-established baseline. 

This choice allows for a fair and meaningful comparison for the application of \textit{TabPFN} \citep{hollmann_2025_tabpfn}, ensuring that improvements are due to the model's capabilities rather than excessive model tuning or architecture complexity. 

\subsubsection{A State-of-the-art foundation model: tabular prior-data fitted network (\textit{TabPFN})}

A foundation model is a deep learning model pre-trained on large datasets, learning general patterns, relationships, and feature interactions across diverse tasks. Unlike traditional models, foundation models can be applied to new, unseen datasets with little or no additional training, a property known as \textit{zero-shot} learning.

\textit{TabPFN} \citep{hollmann_2025_tabpfn} is built on a transformer architecture, originally developed for natural language processing tasks. Transformers use an attention mechanism, which allows the model to focus on important parts of the input data and capture complex relationships between features. In \textit{TabPFN}, this attention mechanism helps assess the relevance of each feature, allowing the model to learn which features are most important.
This approach is more flexible and powerful than traditional decision tree-based models, as it can uncover subtle feature interactions that may be missed by simpler models.

One of the main advantages of \textit{TabPFN} is its ability to perform well without requiring further fine-tuning on specific datasets. This makes it particularly well-suited for small to medium-sized datasets, such as the 4LAC-DR3 (3814 sources). Thanks to its pre-training, \textit{TabPFN} can make accurate predictions on new data without the need for additional training, thus enabling faster deployment and easier integration with multiple catalogues.

Moreover, \textit{TabPFN} processes the entire dataset in a single pass during inference, enabling near-instantaneous predictions. Interestingly, during the "fit" phase, where new training data are presented to the model, it does not optimize its internal weights but rather integrates the given dataset to compute a posterior predictive distribution.
Unlike traditional models, \textit{TabPFN} natively handles missing values by encoding them as meaningful inputs rather than requiring explicit imputation. 
This is achieved through two key mechanisms. First, ordinal encoders allow missing values to propagate through the network, preserving their informational content. Second, \textit{TabPFN}'s pre-training on a diverse set of synthetic datasets, with varying probabilistic relationships between features and target classes, enables it to approximate an optimal probabilistic model at inference time, inherently handling missing values without explicit imputation.

Thanks to its efficient inference process and its robust handling of missing values, \textit{TabPFN} provides a practical and scalable solution for blazar classification across diverse datasets. While pre-processing missing values may still be beneficial for optimal performance, \textit{TabPFN}'s ability to integrate them directly enhances its adaptability to real-world astronomical data.

\textit{TabPFN} has demonstrated superior performance on real-world datasets, outperforming state-of-the-art models such as \textit{Random Forests} \citep{breiman_2001_random} and \textit{XGBoost}, in terms of accuracy and the area under the curve (AUC), a metric measuring the model's ability to distinguish between different classes, \citep[see][Section 5.2]{hollmann_2023_tabpfn}. 
Given these advantages, \textit{TabPFN} represents a significant advancement in tabular data classification and motivated us to compare its performance with our baseline model, \textit{XGBoost}. Its ability to generalize well to different datasets and its efficiency make it an ideal candidate for classifying blazars of the 4LAC-DR3.

\subsection{From classification results to population study}

\begin{figure*}[t]
\centering
\includegraphics[width=0.9\textwidth]{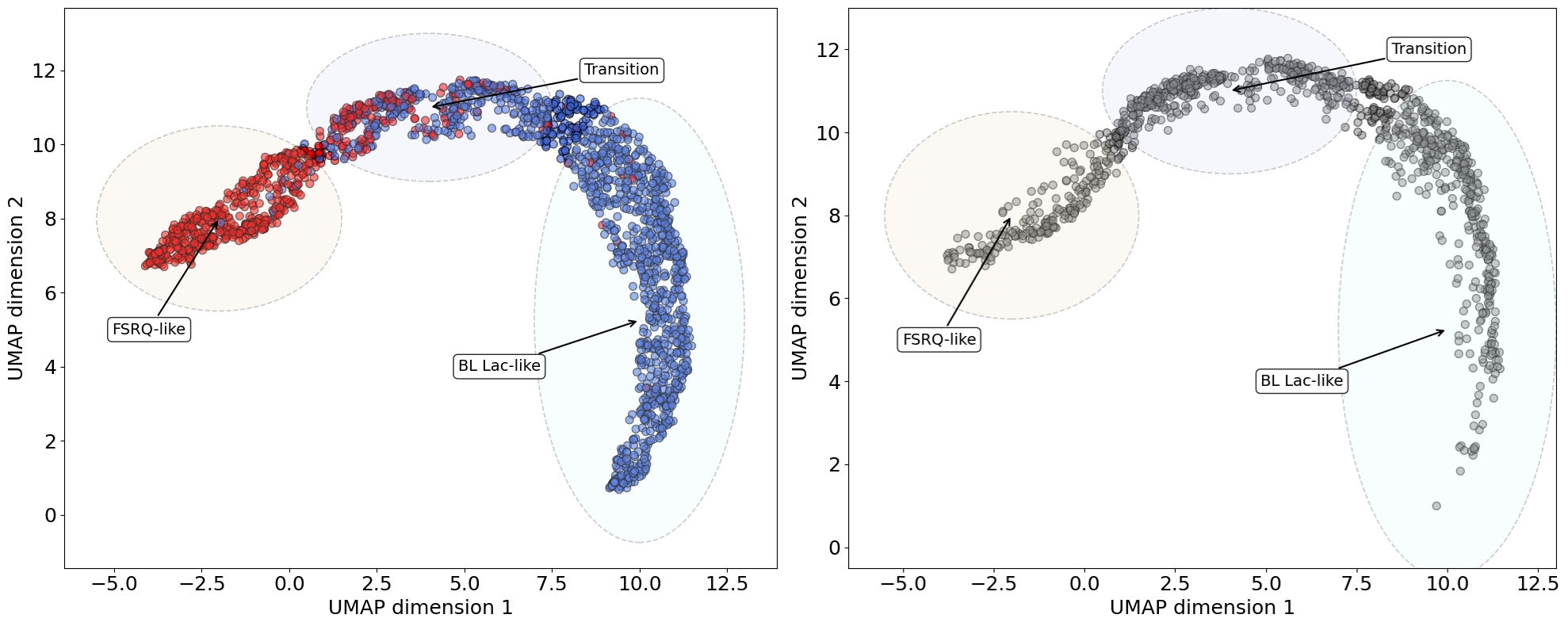}
\caption{Left: Two-dimensional \textit{UMAP} representation of the \textit{TabPFN} latent space for the FSRQs (red) and BL Lacs (blue) of the full FSRQ+BL Lac sample in the 4LAC-DR3 catalogue. Right: Same representation for the BCU sample. To better visualize the regions, ellipses have been overlaid to guide the reader’s eye through the latent space structure.}
\label{fig:latent_space_split}
\end{figure*}

Based on both models, we adopt the following pipeline, where we: 
\begin{itemize}
    \item Split the 4LAC-DR3 dataset into two subsets: the first contains only FSRQs and BL Lacs and is used for optimization and testing; the second consists solely of BCUs and is reserved for subsequent classification.
    \item Perform a feature selection step to retain only the most informative parameters for distinguishing between FSRQs and BL Lacs.
    \item Optimize the parameters of both models.
    \item Evaluate their performance on the FSRQ vs. BL Lac sample using standard classification metrics, and select the best-performing model to infer the class probabilities of BCUs.
\end{itemize}

The complete optimization process is detailed in Appendix \ref{training_opti}, including the dataset splitting strategy (\ref{split_cv}), the selection of features (\ref{FeatureOptimization} and \ref{identify_feat}) and the optimization of the parameter's model (\ref{hyper-param}). The classification performance results are presented in Appendix \ref{results_test}, covering the performance on the FSRQ+BL Lac sample (\ref{fsrq_vs_bll}) and the application to the BCU sample (\ref{application_bcu}). We show that \textit{TabPFN} consistently outperforms \textit{XGBoost} across all tested metrics, which motivates its use as the central model in our analysis.

Importantly, while most sources are confidently classified as either FSRQ or BL Lac, a subset of sources receives more ambiguous predictions. While this ambiguity could stem from model limitations, it might also reflect genuine physical differences, indicating that these sources do not belong strictly to either class. They could instead constitute a separate or transitional population.

To explore this hypothesis, we extract the predicted BL Lac probability, $P_{\mathrm{BLL}}$, for all sources in the sample (FSRQs, BL Lacs, and BCUs). This probability, introduced in Fig.~\ref{fig:distri_BLL_proba} of Appendix \ref{results_test}, and provided by \textit{TabPFN}, quantifies the likelihood that a given source belongs to the BL Lac class. Specifically:
\begin{itemize}
    \item Values close to $P_{\mathrm{BLL}} = 1$ suggest that the source is more likely to be a BL Lac,
    \item Values close to $P_{\mathrm{BLL}} = 0$ suggest that the source is more likely to be an FSRQ,
    \item Intermediate values, centred at $P_{\mathrm{BLL}} = 0.5$ reflect uncertain or ambiguous classifications.
\end{itemize}

Furthermore, the kernel density estimation (KDE) analysis described in Appendix~\ref{fsrq_vs_bll} shows that most BL Lac-like sources exhibit \( P_{\mathrm{BLL}} > 0.9 \), while most FSRQ-like sources fall below \( P_{\mathrm{BLL}} < 0.2 \). Accordingly, we adopt these two thresholds in the following figures to highlight and localize the two populations. Ultimately, using $P_{\mathrm{BLL}}$ rather than discrete labels introduces a continuous dimension to our analysis, allowing for a more nuanced exploration of the blazar population and potential transitional behaviours between types.

Another crucial advantage of using neural network-based models such as \textit{TabPFN} lies in their internal data representation, commonly referred to as the \textit{latent space}. This latent representation encodes the model’s learned features, offering a powerful tool for analysing the global structure of the blazar population. Combined with $P_\mathrm{BLL}$, it enables an in-depth exploration of blazar diversity and potential evolutionary links, as detailed in the following sections.

\section{Results}\label{results}
The goal of this work goes beyond a traditional classification task, as previously explored in various studies \citep{agarwal_2023_classif_blazars, bhatta_2024_classif_blazars, sahakyan_2022_bcu_classification}. We aim to use modern machine learning tools to probe the intrinsic structure of the blazar population, identify potential intermediate sources, and gain insight into their physical nature. In the following sections, we explore the latent space and the associated class probabilities to gain insights into the overall organization of FSRQs, BL Lacs, and BCUs, with particular attention to transitional behaviours and ambiguous sources.

\subsection{BL Lacs versus FSRQs}\label{fsrq_bll_visualization}

Rather than using discrete values to evaluate the model's classification capabilities, an alternative way consists in examining how sources appear within its internal ("latent") representation, or \textit{embeddings}. In the case of \textit{TabPFN}, this latent space is originally high-dimensional (768 dimensions, for \texttt{n\_estimators}$ = 4$, see \ref{hyper-param}), reflecting a learned feature representation that captures the underlying structure necessary to distinguish FSRQs, BL Lacs, and BCUs. By applying the uniform manifold approximation and projection technique (\textit{UMAP}) \citep{McInnes_2018_UMAP} to this high-dimensional space, we can visualize and interpret the distribution of sources in two or three dimensions. This reduced-dimensionality visualization helps assess source separations, identify potential overlaps or classification continua involving BCUs, and detect any unusual or potentially mislabelled sources.

However, before applying \textit{UMAP} to the \textit{TabPFN} embeddings, we first perform a dimensionality reduction using a linear dimensionality reduction technique, called principal component analysis or \textit{PCA}, which reduces the embedding dimension from 768 to 50. This preliminary \textit{PCA} step is essential because \textit{UMAP} exhibits quadratic computational complexity, making it inefficient and potentially unstable when applied directly to high-dimensional data. The application of this method on the sample composed of FSRQs and BL Lacs yields the results shown in the left panel of Fig.~\ref{fig:latent_space_split}. For the \textit{UMAP} projection, we adopt the following hyper-parameters: \texttt{n\_neighbours} = 15, \texttt{metric} = \texttt{euclidean}, and \texttt{min\_dist} = 0.1.

Another important clarification concerns the interpretation of the latent space structure presented in Fig.~\ref{fig:latent_space_split}. Before the application of the \textit{UMAP} projection, the data were processed through a supervised classification model, namely \textit{TabPFN}, trained using the source labels provided in the 4LAC-DR3 catalogue (as detailed in Appendix~\ref{results_test}). As a result, both the predicted probabilities and the intermediate embeddings produced by the model inherently reflect this supervised information. This setup naturally leads to a \textit{UMAP} projection in which the latent space aligns with the BL Lac probability \( P_{\mathrm{BLL}} \), leading to a smooth and continuous representation. 

It is important to emphasize that the specific geometric shape observed in the \textit{UMAP} projection of the latent space, which, in our case, is an arch-like structure, does not carry intrinsic physical meaning. This layout is a product of the \textit{UMAP} algorithm’s initialization, and different parameter choices yield differently-shaped projections. What is meaningful, however, is the smooth correlation between the layout and the observational properties of the sources, such as the spectral index, the flux, the variability, etc. 
This correlation guarantees that our conclusions are driven by meaningful associations, rather than artifacts of the projection technique.

Looking through this continuous structure we observe an elongated region on the right, predominantly composed of BL Lacs (blue points), while FSRQs (red points) are mostly concentrated on the left. Between the BL-Lac dominated region (right) and the FSRQ-dominated region (left), we identify a third, intermediate region that can be distinguished from the two main populations. The dominant source type in this region is less clearly defined, and it may contain objects whose properties slightly deviate from both FSRQs and BL Lacs. These sources could potentially represent a transitional population between the two classes. A more comprehensive discussion on this transitional behaviour, incorporating all source types, is provided in Section \ref{discussion}. Interestingly, we observe that some FSRQs are located in the region dominated by BL Lacs and vice versa. This likely corresponds to originally misclassified sources in the catalogue. Finally, it is important to note that in Fig.~\ref{fig:latent_space_split}, the \textit{UMAP} dimensions (X and Y axes) do not carry direct physical meaning, as they only reflect the coordinates of each source in the reduced-dimensional space.

Finally, we fix the two-dimensional \textit{UMAP} mapping fitted on the 50-dimensional feature space of the FSRQs and BL Lacs, and apply this same transformation to all subsequent visualizations. Freezing this \textit{UMAP} non-linear projection preserves a consistent relationship between the latent space and its two-dimensional representation, ensuring that all plots share the same geometric mapping. This consistency enables a reliable interpretation of the distribution of the sources in the latent space and facilitates a direct comparison between different blazar subsets.

\subsection{The BCU population}\label{bcu_visualization}

At this step, we can visualize in the right panel of Fig.~\ref{fig:latent_space_split} the distribution of BCUs in the same latent space representation presented in Section \ref{fsrq_bll_visualization}.
By comparing this to the left panel, where FSRQs and BL Lacs are found in opposite regions of the latent space, we observe that some BCUs fall within the regions dominated by either FSRQs (left) or BL Lacs (right). However, a significant fraction lies in the intermediate zone, previously discussed in Section \ref{fsrq_bll_visualization}, where the dominant class is not well defined. This spatial overlap suggests a dual nature of the BCU population: some can be reliably associated with one of the two known classes, while others may possess hybrid or atypical characteristics that prevent a straightforward classification. This indicates that some BCUs exhibit properties shared between FSRQs and BL Lacs, making their classification more challenging. Such intermediate characteristics could point to a transitional population, a possibility we examine more closely in the next section.

\subsection{Population study}\label{population_study}

\begin{figure}[t]
\centering
\includegraphics[width=0.5\textwidth]{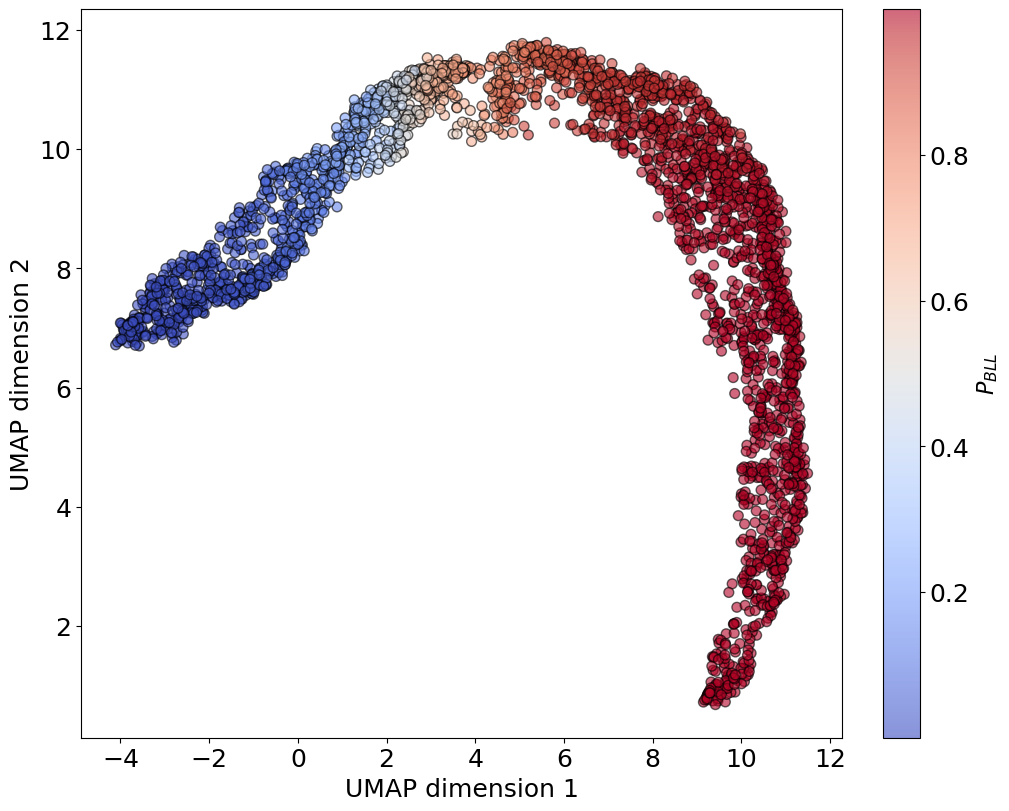}
\caption{Two-dimensional \textit{UMAP} representation of the \textit{TabPFN} latent space for the (FSRQ+BL Lac+BCU) blazar sample coloured by the $BLL$ probability $P_{\mathrm{BLL}}$.}
\label{fig:latent_space_pbll_4lac}
\end{figure}

Once a probability $P_{\mathrm{BLL}}$ (along with its associated uncertainty) has been obtained for all sources, we can use the \textit{UMAP}-reduced dimensional latent space representation of \textit{TabPFN} as a pivotal tool to explore the properties of the entire blazar population. This approach not only enables us to determine where BCUs lie in relation to FSRQs and BL Lacs, but also to characterize all sources in our sample, thereby gaining deeper insight into potential links between different types of blazars. To this end, we assign a BL Lac probability to all sources in the 4LAC-DR3.

We first visualize the latent space for the full blazar sample, comprising FSRQs, BL Lacs, and BCUs, colour-coded by their respective $P_{\mathrm{BLL}}$ values. This representation allows us to evaluate how well the model captures a continuum between the two main classes. As shown in Fig.~\ref{fig:latent_space_pbll_4lac}, the two extremes are well-separated, with FSRQ-like sources having $P_{\mathrm{BLL}} < 0.2$ and BL Lac-like sources with $P_{\mathrm{BLL}} > 0.9$. An intermediate population fall between these, within the interval $0.2 \leq P_{\mathrm{BLL}} \leq 0.9$.

\begin{figure}[t]
\centering
\includegraphics[width=0.5\textwidth]{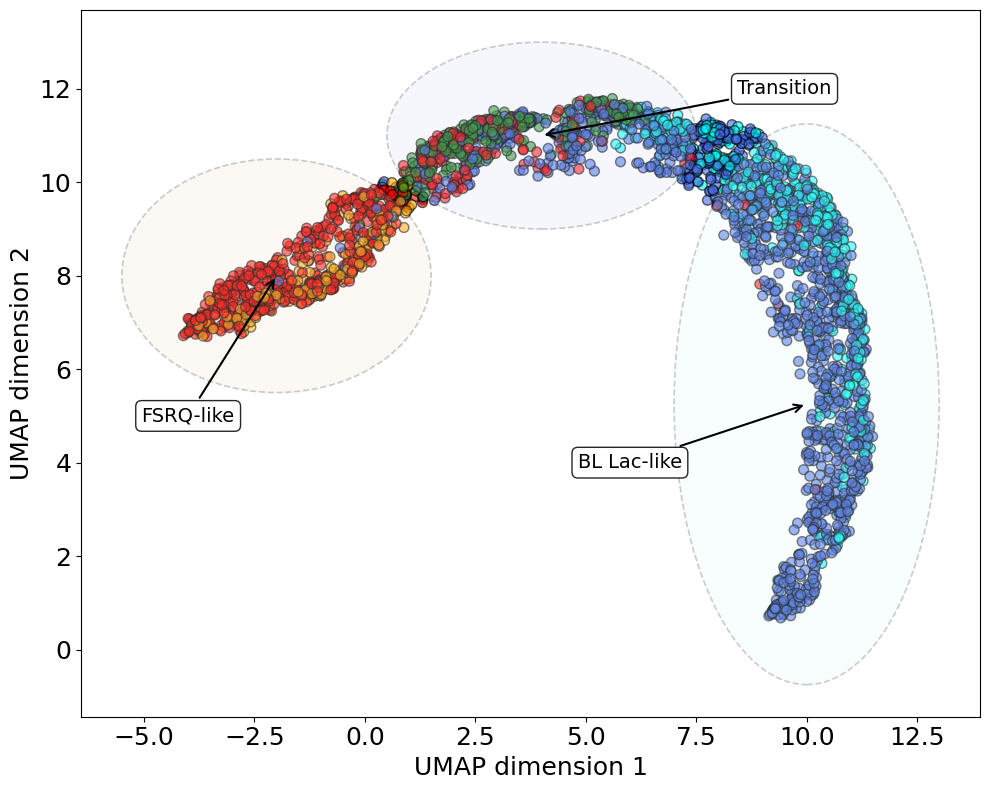}
\caption{Two-dimensional \textit{UMAP} representation of the \textit{TabPFN} latent space for the full blazar sample including FSRQs (red), FSRQ-like BCUs with $P_{\mathrm{BLL}}<0.2$ (orange), BL Lacs (dark blue), BL Lac-like BCUs with $P_{\mathrm{BLL}}>0.9$ (light blue), and BCUs with $0.2\leq P_{\mathrm{BLL}}\leq 0.9$ (green) for the full blazar sample.}
\label{fig:latent_space_all_4lac}
\end{figure}

We then display the same latent space, this time coloured according to the source type (Fig.~\ref{fig:latent_space_all_4lac}).  Similarly to the left panel of Fig.~\ref{fig:latent_space_split}, we identify on the left the region dominated by FSRQ-like sources. When considering the full blazar sample, this includes both Fermi-classified FSRQs (red) and BCUs with $P_{\mathrm{BLL}} < 0.2$ (orange). On the right, we retrieve the region dominated by BL Lac-like sources, consisting of Fermi BL Lacs (dark blue) and BCUs with $P_{\mathrm{BLL}} > 0.9$ (light blue). In contrast, sources falling within the probability range $0.2\leq P_{\mathrm{BLL}}\leq 0.9$ and located in the intermediate region are a mix of FSRQs, BL Lacs, and BCUs.  This suggests a heterogeneous group whose properties blend those of FSRQs and BL Lacs.

This visualization indicates that the conventional FSRQ vs. BL Lac classification holds primarily for extreme cases (e.g., HSP BL Lacs vs. FSRQs), but may require refinement in the intermediate region, where objects with mixed or evolving properties dominate. The presence of this intermediate region supports the concept of CLBs, as proposed in recent studies \citep[e.g.,][]{kang_2024_clb}.

\begin{figure*}[t]
\centering
\includegraphics[width=1\textwidth]{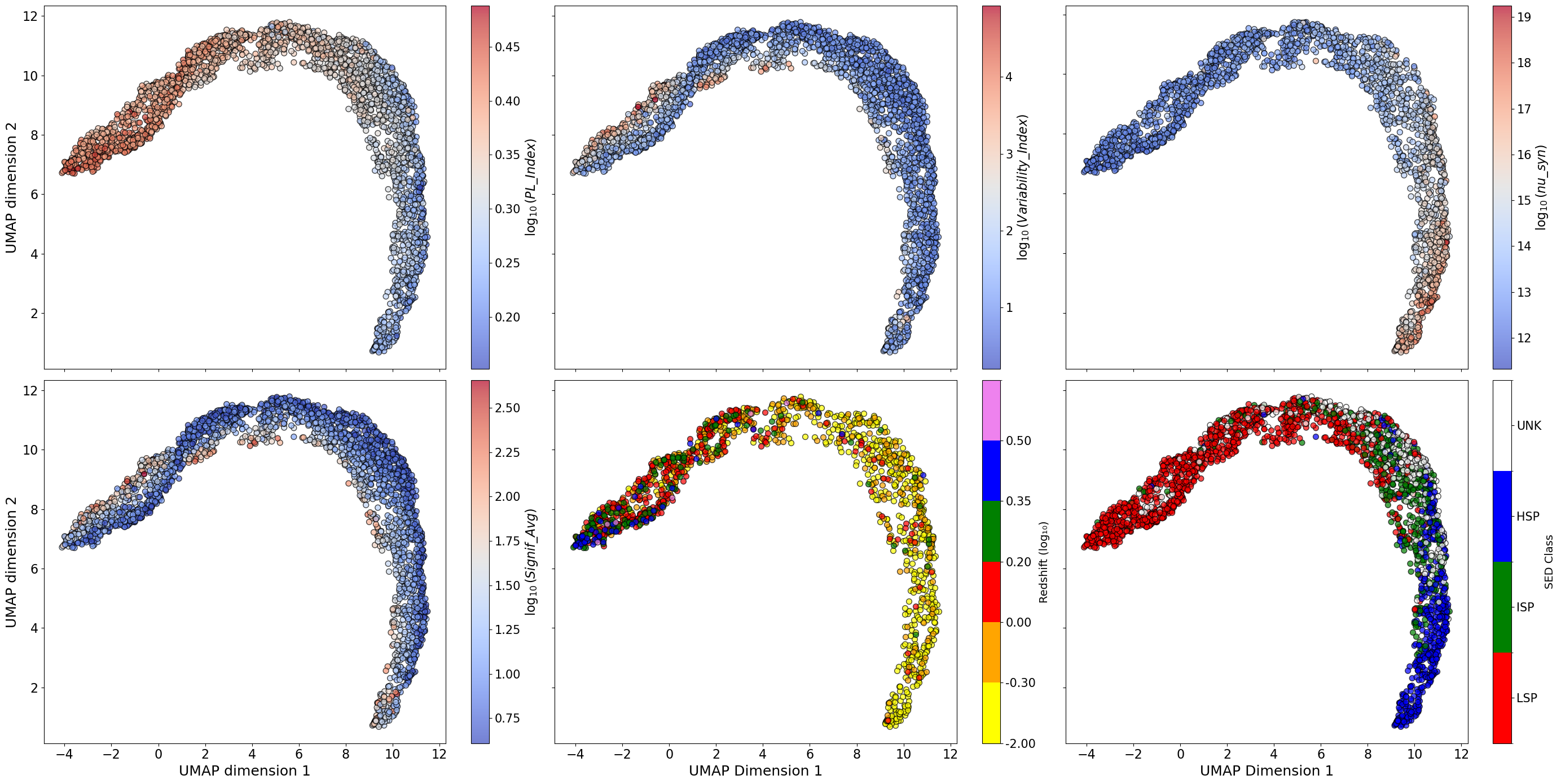}
\caption{Two-dimensional \textit{UMAP} representation of the latent space for the full 4LAC-DR3 sample, coloured with the high-energy photon index \texttt{PL\_Index} (top left), \texttt{Variability\_Index} (top middle) the frequency of the synchrotron peak \texttt{nu\_syn} (top right), the detection significance \texttt{Signif\_Avg} (bottom left), the redshift \texttt{z} (bottom middle) and the \texttt{\texttt{SED\_class}} (bottom right). For better visibility, missing values of the redshift (\texttt{z}) and of the synchrotron peak (\texttt{nu\_syn}) are not displayed.}
\label{fig:latent_space_gammainfo}
\end{figure*}

It is important to emphasize that our study focuses on a slow, long-term evolutionary scenario, consistent with the interpretation proposed by \citet{kang_2024_clb}. In this framework, CLBs correspond to FSRQs that gradually evolve toward a BL Lac state through a progressive decrease in accretion efficiency, potentially entering the ADAF regime over cosmological timescales. As such, our definition of CLBs excludes cases like OQ~334, which are characterized by rapid and reversible state changes on timescales of days to months (as previously discussed in the introduction). Accordingly, throughout this work, we adopt the terminology of \citet{kang_2024_clb} and refer to the sources located in the intermediate region of our latent space as CLBs.

To deepen our understanding of this transition, we next provide complementary visualizations of the same latent space, in which points are colour-coded by additional physical properties. We first use the \textit{SED} information, which specifies the position of the synchrotron peak when available. As shown at the bottom right of Fig.~\ref{fig:latent_space_gammainfo}, we observe, as expected, that the left region, dominated by FSRQ-like objects, consists primarily of LSP sources. On the right, it is clear that HBLs occupy the lower part of the BL Lac region, with a transition through IBLs to LBLs in the upper part. Interestingly, in the transition region, the CLBs tend to be LSP-like, which aligns with their intermediate position between FSRQs and LBLs in Fig.\ref{fig:latent_space_all_4lac}. Then, we use the features that were most important for discriminating FSRQs from BL Lacs in our model. As shown in Fig.~\ref{fig:4lac_feat_importance}, these include the synchrotron peak frequency (\texttt{nu\_syn}) and the high-energy photon index (\texttt{PL\_Index}), both of which are closely tied to the physical nature of the sources. The reduced latent space coloured by these features is presented in the top left and top right panels of Fig.~\ref{fig:latent_space_gammainfo}. These visualizations confirm key trends :
\begin{itemize}
    \item FSRQ-like sources are located in a region with low $\log_{10}(\texttt{nu\_syn})\!<\!13$ and soft spectra ($\texttt{PL\_Index}\!>\!2.4$).
    \item Conversely, BL Lac-like sources are found in areas with high $\log_{10}(\texttt{nu\_syn})\!>\!16$ and hard spectra ($\texttt{PL\_Index}\!<\!2$).
    \item CLBs occupy a transitional zone, with intermediate values: $13\!<\!\log_{10}(\texttt{nu\_syn})\!<\!15$ and $2\!<\!\texttt{PL\_Index}\!<\!2.4$.

\end{itemize}

\begin{figure*}[t]
\centering
\includegraphics[width=0.9\textwidth]{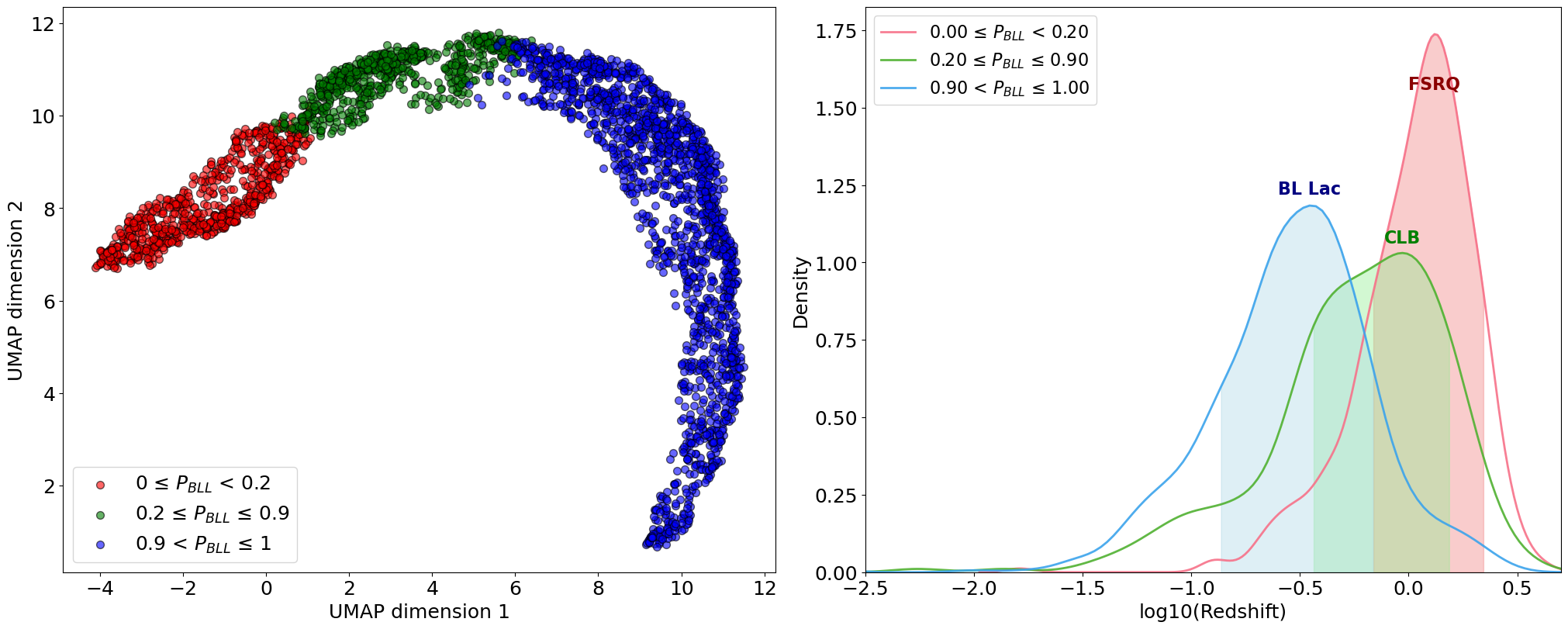}
\caption{Left: Two-dimensional \textit{UMAP} representation of the latent space for the full 4LAC-DR3 sample split into $P_{\mathrm{BLL}}$ bins. Right: corresponding redshift density distributions. Shaded regions represent high-density intervals based on a KDE}.
\label{fig:population_z}
\end{figure*}

It is important to emphasize that the intermediate values of the photon index \texttt{PL\_Index} observed for CLBs are not an artifact of limited detection significance. As shown in the bottom left panel of Fig.~\ref{fig:latent_space_gammainfo}, the average detection significance (\texttt{Signif\_Avg}) appears to be smoothly distributed across the latent space, without being confined to a specific region. This suggests that the intermediate \texttt{PL\_Index} values found in the transition region genuinely reflect intrinsic physical properties of these sources, rather than being driven by observational biases.

Interestingly, we also identify a distinct branch in the latent space associated with sources exhibiting higher \texttt{Signif\_Avg} values. As shown in the top middle panel of Fig.~\ref{fig:latent_space_gammainfo}, this branch corresponds to highly variable sources. By examining Fig.~\ref{fig:latent_space_all_4lac}, we note that this region is primarily populated by non-BCU sources, suggesting that BCUs in the 4LAC-DR3 generally exhibit lower levels of variability.

Another key feature for investigating source properties is the redshift \texttt{z}. As mentioned in Section \ref{4LAC-DR3}, FSRQs are generally found at higher redshift than BL Lacs, a trend well documented in recent blazar catalogues like the 4LAC-DR3 \mbox{\citep{ajello_2022_4lac}}. When colouring the latent space with the redshift (bottom middle panel of Fig.~\ref{fig:latent_space_gammainfo}), this trend is indeed recovered: high-redshift sources are concentrated on the left region (dominated by FSRQs), while low-redshift sources are found on the right (dominated by BL Lacs). However, the transition between these regions is less evident, likely due to incomplete redshift data making it difficult to fully resolve a continuous distribution.

To further investigate this transition and better characterize the redshift distribution of each population, we divide the continuous $P_{\mathrm{BLL}}$ distribution into three intervals, using the thresholds defined in Appendix~\ref{fsrq_vs_bll}:
(i) $P_{\mathrm{BLL}} < 0.2$,
(ii) $0.2 \leq P_{\mathrm{BLL}} \leq 0.9$, and
(iii) $P_{\mathrm{BLL}} > 0.9$.
The goal is to assess whether we can distinguish three distinct groups: high-redshift FSRQs, low-redshift BL Lacs, and an intermediate group potentially corresponding to CLBs. The results are shown in Fig.~\ref{fig:population_z} using the KDE algorithm.

It is important to note that the shaded regions do not represent formal confidence intervals. Instead, they correspond to high-density intervals centred around the peak values of the distributions, as defined using a KDE approach. This method is well suited to account for potentially broad or asymmetric distribution shapes, and it helps to better visualize and interpret the results. The redshift distributions indeed reveal three distinguishable populations:
\begin{itemize}
    \item FSRQs ($P_{\mathrm{BLL}} < 0.2$): mean redshift $\langle \texttt{z} \rangle = 1.25$, with a high-density interval of $[0.59, 1.90]$.
    \item BL Lacs ($P_{\mathrm{BLL}} > 0.9$): $\langle \texttt{z} \rangle = 0.40$, with $[0.03, 0.77]$.
    \item CLBs ($0.2 \leq P_{\mathrm{BLL}} \leq 0.9$): $\langle \texttt{z} \rangle = 0.79$, with $[0.20, 1.40]$.
\end{itemize}

While CLBs do occupy an intermediate range of redshift, their distribution is flatter and overlaps with both the FSRQ and BL Lac distributions, making them less sharply defined. Nonetheless, this result aligns with the findings of \citet{kang_2024_clb}, where CLBs also appear at intermediate redshift, supporting the hypothesis of a cosmological evolution from FSRQs to BL Lacs.

\begin{figure*}[ht]
    \centering
    \includegraphics[width=0.85\textwidth]{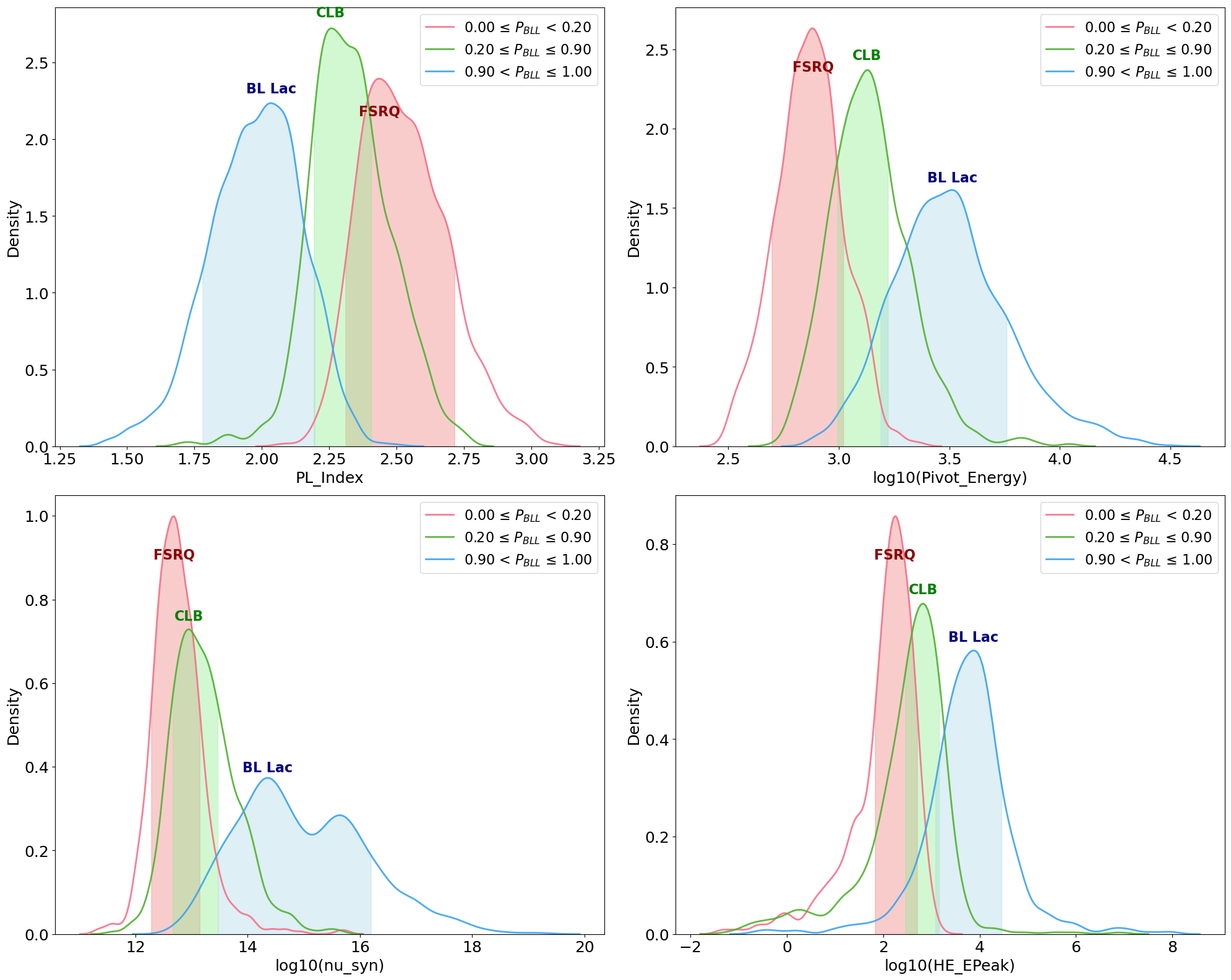}
    \caption{\texttt{PL\_Index}, \texttt{Pivot\_energy}, \texttt{nu\_syn} and \texttt{HE\_EPeak} density distributions for different $P_{\mathrm{BLL}}$ bins.}
    \label{fig:transition_features}
\end{figure*}

Furthermore, their study reported a similar transition in the photon index, moving from soft FSRQ spectra to hard BL Lac spectra through intermediate values for CLBs. This trend is consistent with the 4LAC-DR3 results \citep[see][Fig. 1]{ajello_2022_4lac} where three distinct populations of photon index (corresponding to FSRQs, BCUs, and BL Lacs) were noted. We expand on this aspect in Fig.~\ref{fig:transition_features}, where we show the density distributions of \texttt{PL\_Index} and other meaningful features across the three $P_{\mathrm{BLL}}$ bins. As expected, since \texttt{PL\_Index} is the most discriminative feature for the model, we observe in the top left panel of Fig.~\ref{fig:transition_features} three well-separated populations in terms of photon index, with a value of \texttt{PL\_Index} centred around $2.3$ for CLBs. Similar patterns appear in the distributions of \texttt{Pivot\_energy} (top right panel) and \texttt{HE\_EPeak} (bottom right panel). Although the distinction in \texttt{nu\_syn} (bottom left panel) is less pronounced, likely due to the high rate of missing data, the overall trend remains consistent: CLBs exhibit intermediate values between FSRQs and BL Lacs.

These results, together with the gradual evolution of the source classification based on the SED properties, as illustrated in the bottom right panel of Fig.~\ref{fig:latent_space_gammainfo}, support a scenario in which CLBs represent an intermediate evolutionary phase within a continuous transition from FSRQs to BL Lac objects. In this transition, sources progress through the sequence: FSRQ~$\rightarrow$~CLB~$\rightarrow$~LSP BL Lac~$\rightarrow$~ISP BL Lac~$\rightarrow$~HSP BL Lac, reflecting a gradual shift from low- to high-synchrotron-peaked spectral energy distributions.
Their physical properties, across both SED features and gamma-ray spectra, suggest that their accretion disks are not yet fully depleted, but are evolving toward a radiatively inefficient state. The transition we find is consistent with evolutionary models of blazars where the gradual depletion of the circumnuclear material leads to changes in the accretion regime \citep{cavaliere_2002_blazar, boettcher_2002_blazar_evolution}. In the early stages, FSRQs are fuelled by gas-rich environments that allow high accretion rates and luminous, radiatively efficient disks \citep{yuan_2014_hot_accretion, prandini_2022_blazar_sequence}. Over cosmic time, the reservoir of gas diminishes, accretion weakens, and the sources transition into the BL Lac population. In this framework, the sources in the transition region of the latent space are likely undergoing this shift, from efficient to inefficient accretion. This scenario naturally explains the continuous evolution of observational properties such as spectral indices, emission line strengths, and SED peak locations. If the depletion is driven by AGN feedback mechanisms or galaxy mergers, these processes could also regulate the timescales of the transition \citep{fabian_2012_agn_feedback}.

\subsection{Towards a cosmological evolution: the central engine}\label{cosmological_evolution}

Finally, in the context of the evolutionary scenario proposed by \citet{cavaliere_2002_blazar} and \citet{boettcher_2002_blazar_evolution}, if FSRQs evolve towards a radiatively inefficient accretion regime, characteristic of BL Lacs, such evolution should also be reflected in the physical properties of the central engine.

To investigate this aspect, \citet{kang_2024_clb} made use of the catalogue of central engine properties for Fermi blazars compiled by \citet{paliya_2021_central_engines}. As mentioned in Section \ref{4LAC-DR3}, this catalogue provides key parameters such as the black hole mass ($M_{\mathrm{BH}}$), the logarithm of the accretion disk luminosity (\texttt{logLd}) and the Compton dominance (\texttt{CD}), defined as the ratio between the inverse Compton and synchrotron peak luminosities, an indicator of the relative strength of the high-energy emission. Their analysis reveals that CLBs lie in an intermediate region of the feature space concerning both \texttt{logLd} and \texttt{CD}, supporting the idea of a gradual transition in accretion properties. However, no such trend was found for $M_{\mathrm{BH}}$, suggesting that black hole mass alone may not play a primary role in driving this evolution.

\begin{figure*}[t]
\centering
\includegraphics[width=1\textwidth]{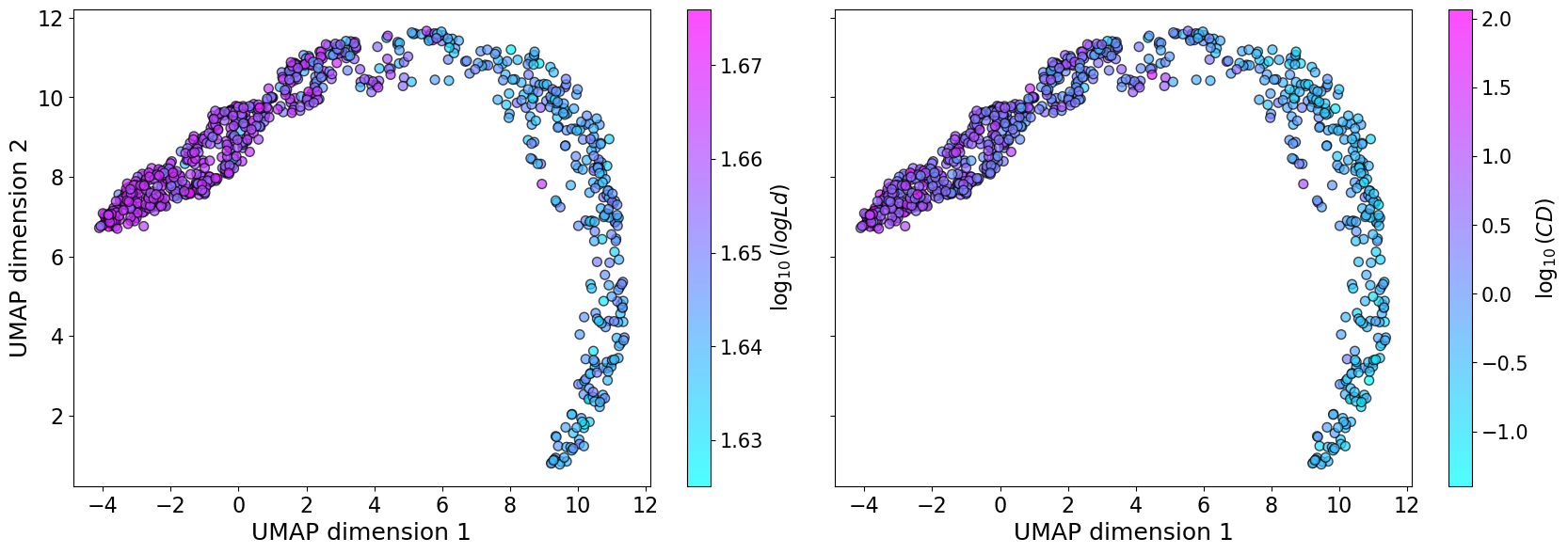}
\caption{Two-dimensional latent space representation of the full blazar sample, coloured with \texttt{logLd} (left panel) and \texttt{CD} (right panel). Missing values are not shown.}
\label{fig:latent_space_BHinfo}
\end{figure*}

\begin{figure*}[t]
    \centering
    \includegraphics[width=0.9\textwidth]{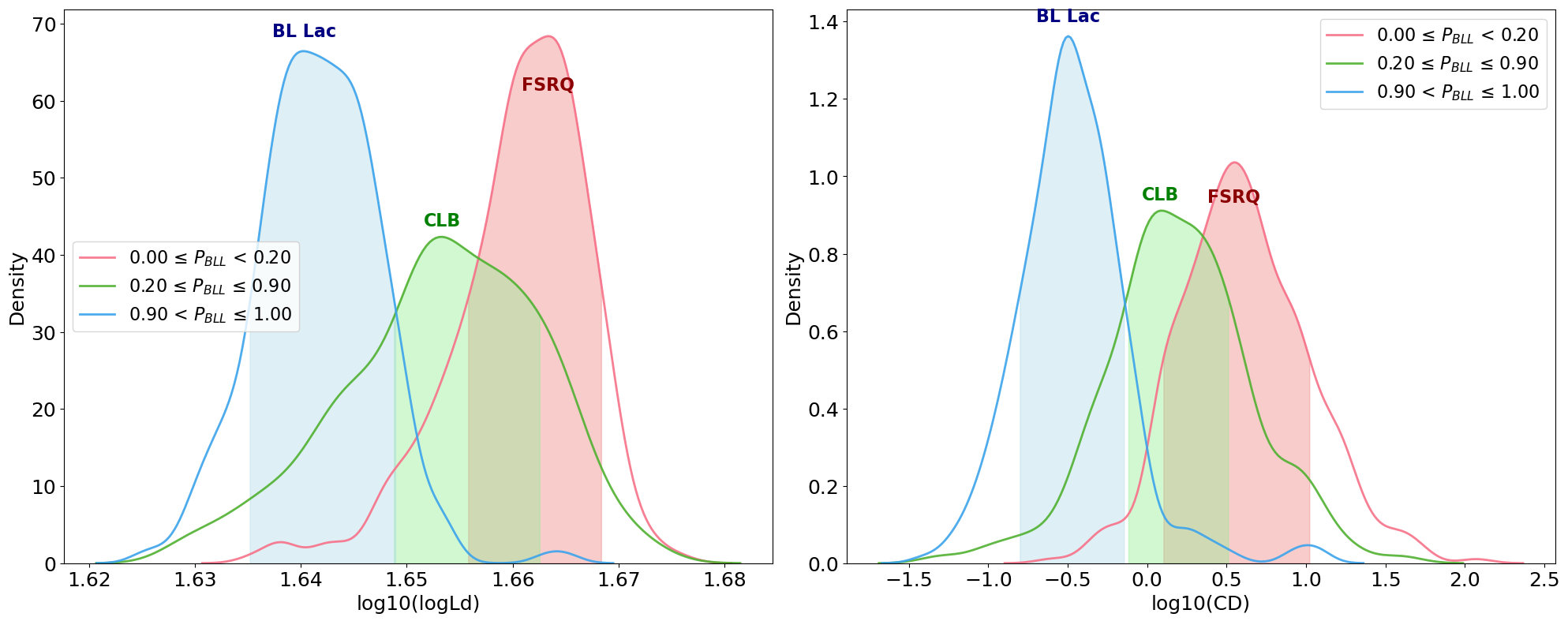}
    \caption{\texttt{logLd} and \texttt{CD} density distributions for different $P_{\mathrm{BLL}}$ bins.}
    \label{fig:population_BH}
\end{figure*}

Building on these results, we present in Fig.~\ref{fig:latent_space_BHinfo} our reduced latent space representation coloured by $\texttt{logLd}$ and $\texttt{CD}$.
Additionally, we further apply a logarithmic transformation to \texttt{logLd} and \texttt{CD} to better highlight differences between populations in the reduced latent space. Prior to this transformation, we ensured that only strictly positive values of \texttt{logLd} and \texttt{CD} were retained, to avoid undefined or non-physical values. These maps clearly show a distinction between FSRQs, characterized by high \texttt{logLd} and \texttt{CD} values on the left side of the latent space, and BL Lacs, which exhibit significantly lower values and dominate the right region. This is consistent with the physical interpretation that FSRQs harbour radiatively efficient accretion disks and strong external photon fields, resulting in a more prominent inverse Compton component. We note that in the FSRQ-dominated region of the latent space, a few sources show lower-than-average values of CD and logLd. This may reflect a transitional phase, where the accretion disk and external photon fields are fading, reducing both observables while the source still displays broad optical lines and is thus classified as an FSRQ. Misclassification may also contribute: as shown in Fig.~\ref{fig:distri_BLL_proba}, a small fraction of BL Lacs in the 4LAC catalogue have a low BL Lac probability ($P_{\mathrm{BLL}} < 0.2$) and fall within the FSRQ region. These represent only about 1\% of the total, and thus have minimal impact on the overall interpretation. As in previous sections, we further investigate this transition by displaying the density distributions of \texttt{logLd} and \texttt{CD} across the three $P_{\mathrm{BLL}}$ bins, as shown in Fig.~\ref{fig:population_BH}. These distributions again reveal a continuous transition between FSRQs and BL Lacs, with CLBs occupying the intermediate regime. To better visualize the central trend, we also report the mean of \texttt{logLd} and \texttt{CD} in logarithmic units. The resulting average values in log scale are $\langle\log_{10}( \texttt{logLd} )\rangle = 1.65$ and $\langle\log_{10}( \texttt{CD} )\rangle = 0.44$, reinforcing the classification of CLBs as a transitional population in terms of central engine activity.

Together with the results in Section \ref{population_study}, which showed similar transitions in redshift and photon index, these findings further support a scenario in which blazars evolve from FSRQs to BL Lacs via an intermediate CLB phase. Initially, young FSRQs possess high accretion rates and standard thin accretion disks, producing strong emission lines. Over time, as their gas reservoir depletes, they transition into CLBs with intermediate accretion properties, eventually becoming BL Lacs, whose central engines are characterized by low accretion rates and radiatively inefficient flows, such as ADAFs \citep{yuan_2014_hot_accretion, prandini_2022_blazar_sequence}.

In this work, we not only confirm the trends reported in \citet{kang_2024_clb}, but also introduce a unified latent representation that allows for a direct and intuitive visualization of the blazar evolutionary sequence. This latent space framework, combined with the inferred probability $P_{\mathrm{BLL}}$, provides a powerful tool to track the progressive changes in both jet and central engine properties across the blazar population.

\section{Discussion}\label{discussion}
\subsection{A new way of characterizing blazars}

Based on these results and our latent space representation, we propose an alternative framework for characterizing blazars. Instead of assigning a discrete class label, each source can be described using two complementary and continuous descriptors: \begin{itemize} 
\item Its probability $P_{\mathrm{BLL}}$ of being a BL Lac. 
\item Its position in the \textit{UMAP}-projected latent space of the \textit{TabPFN} model. \end{itemize}
By combining these two pieces of information, we move beyond rigid binary classification and enable a more flexible, interpretable, and physically motivated description of the blazar population. This approach is particularly well suited for sources with ambiguous or intermediate properties, such as BCUs or, more generally, CLBs, and can be directly applied to future releases of blazar catalogues. 

\begin{figure}[t]
    \centering
    \includegraphics[width=0.5\textwidth]{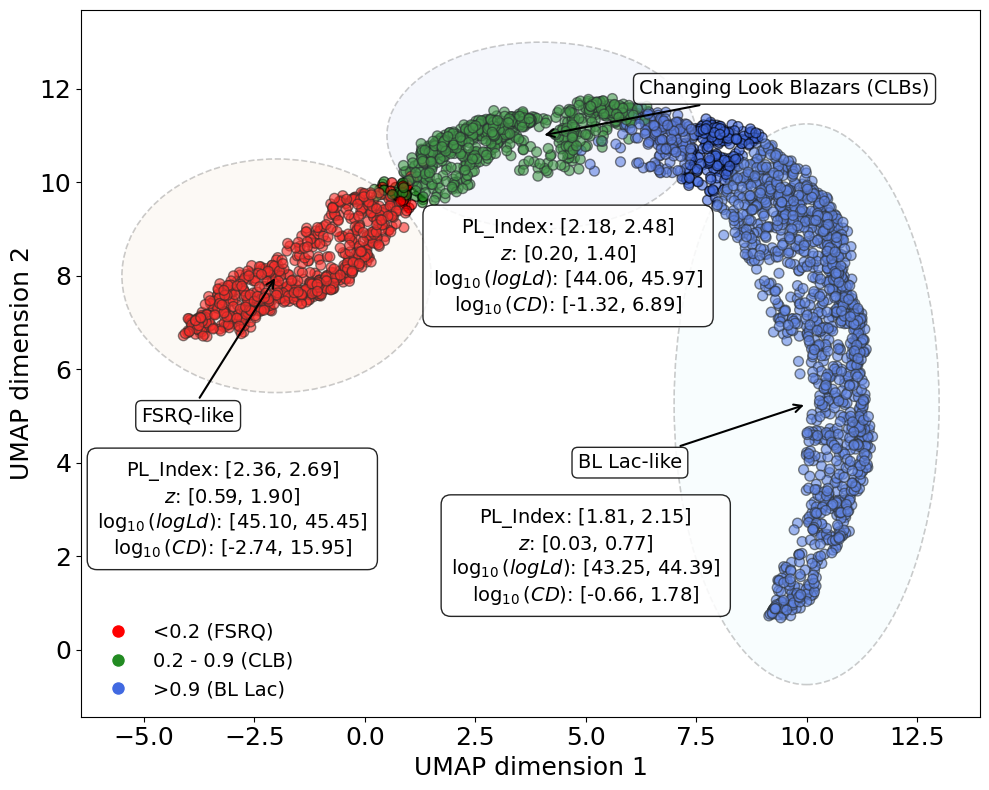}
    \caption{Unified representation of blazar evolution. The continuous \textit{UMAP} representation of the \textit{TabPFN} latent space has been divided into three bins of $P_{\mathrm{BLL}}$, corresponding to the three potential evolutionary phases. For each phase, we provide characteristic ranges for the most discriminating features. These ranges represent high-density intervals around the KDE peaks identified in the histograms presented in previous sections.}
    \label{fig:unified_scheme_blazar}
\end{figure}

Furthermore, this framework allows for an intuitive exploration of blazar evolution. By colour-coding the latent space with key features (e.g., spectral indices, accretion-related quantities), users can easily investigate the progressive transition from FSRQs to BL Lacs, with CLBs forming an intermediate bridge. This visual exploration can be enriched with histograms (see Section~\ref{cosmological_evolution}) to define typical parameter ranges across the transition.
In light of these findings, we propose a unified visualization of blazar evolution, shown in Fig.~\ref{fig:unified_scheme_blazar}. Each evolutionary phase, discretized into three bins of $P_{\mathrm{BLL}}$, is associated with characteristic ranges for the most discriminating features.  These include the gamma-ray photon index, redshift, accretion disk luminosity, and Compton dominance, which act as proxies for the underlying physical processes driving the transition from FSRQs to BL Lacs. This representation constitutes a significant step towards a unified scheme of blazar evolution. It offers a data-driven and interpretable framework for the understanding of the diversity in the blazar population and their potential evolutionary connection.

Finally, to strengthen our results, we include a sample of CLB candidates identified in \citet{Kang2025_CLB_Mclust}. In that study, the authors built upon their previous work \citep{kang_2024_clb}, where CLBs were defined as sources showing intermediate values in key physical parameters such as the Compton dominance, the accretion disk luminosity, and the redshift.

In their follow-up analysis, \citet{Kang2025_CLB_Mclust} applied an unsupervised clustering algorithm to these parameters and identified four optimal combinations that could effectively separate FSRQs, BL Lacs, and CLBs. They compiled a catalogue that includes the original classification from \citet{kang_2024_clb} alongside the clustering results for the four parameter combinations.

\begin{figure}[t]
    \centering
    \includegraphics[width=0.496\textwidth]{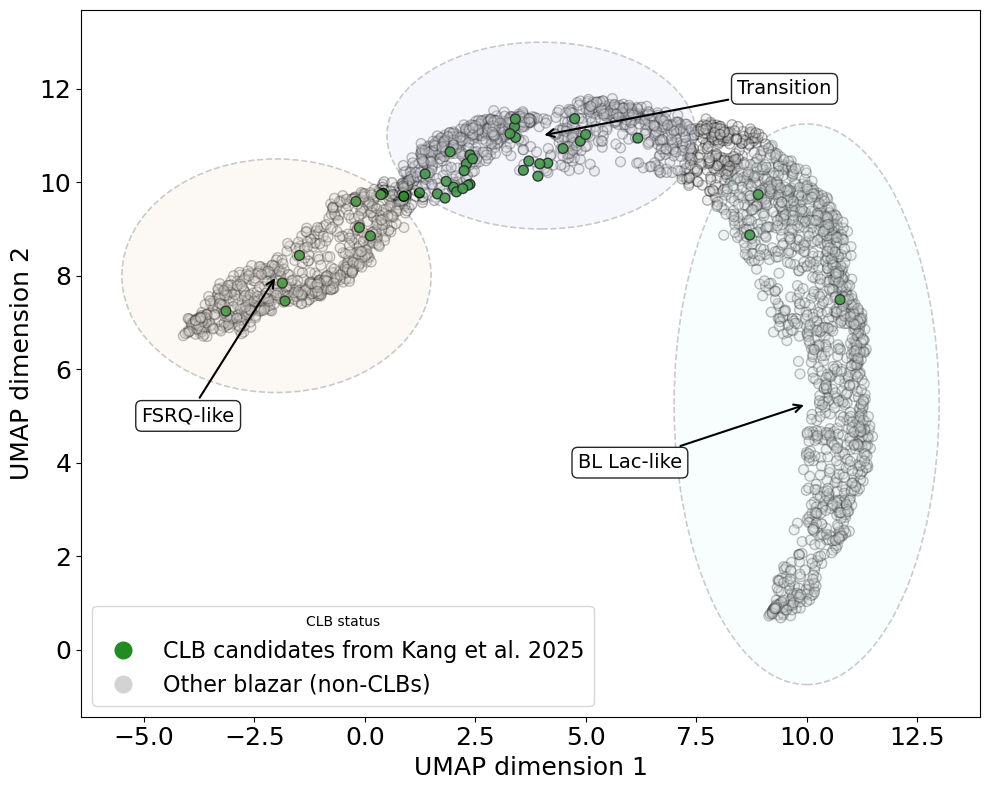}
    \caption{
    Two-dimensional \textit{UMAP} representation of the \textit{TabPFN} latent space showing a subset of CLBCs from \citet{Kang2025_CLB_Mclust}, highlighted in green. All other sources in the sample are shown in grey. The clustering of CLBCs in the intermediate region supports the interpretation of this zone as a transitional phase between FSRQ-like and BL Lac-like states.}
    \label{fig:clb_kang}
\end{figure}

To ensure consistency with the gradual evolutionary scenario proposed in \citet{kang_2024_clb}, we selected only the sources that were: (i) classified as CLBs in the original study, and (ii) consistently identified as CLB candidates (CLBCs) in all four clustering combinations of \citet{Kang2025_CLB_Mclust}. The resulting sample consists of 42 blazars, which we display in our latent space projection (Fig.~\ref{fig:clb_kang}).

We observe that the majority of CLBCs from \citet{kang_2024_clb, Kang2025_CLB_Mclust} fall within, or very close to, the intermediate region of our reduced latent space. This observation supports our interpretation that this region corresponds to a transitional phase, where sources evolve gradually from an FSRQ-like state toward a BL Lac-like state.

A smaller fraction of CLBCs are found outside this region, primarily in the area dominated by FSRQs. This is likely due to the overlap between the CLB and FSRQ clusters in the feature space of \citet{Kang2025_CLB_Mclust} (see Fig. 5 therein), indicating that some sources classified as CLBs still share characteristics with the FSRQ population.

Overall, the spatial consistency between our latent space representation and the independently identified CLBCs provides strong support for the validity of our approach. It confirms the ability of the latent representation to reflect meaningful evolutionary trends within the blazar population, and demonstrates its potential as a powerful framework to identify and investigate sources undergoing, or susceptible to undergo, long-term transitions between blazar subclasses.

\subsection{A tool to investigate missing data}

An additional and particularly compelling application of the reduced latent space representation is its potential to provide insights into missing data. As illustrated at the bottom right of Fig.~\ref{fig:latent_space_gammainfo}, we overlaid the regions where the \texttt{SED\_class} information, available in the 4LAC-DR3 catalogue, is missing. Interestingly, the spatial coherence in the distribution of sources with known SED classes suggests that their neighbours in latent space share similar physical properties.
This implies that the latent space serves as a valuable proxy to infer likely values for missing observational quantities, such as the synchrotron peak position, the redshift, or even the black hole mass. While not a substitute for direct measurements, such inferences can offer useful priors or guide observational follow-ups. In this way, our latent representation becomes a powerful tool not only for classification and population studies but also for mitigating the impact of incomplete data in blazar catalogues.

\section{Conclusions}

In this study, we explored the application of the \textit{TabPFN} model for blazar classification on a major dataset: the 4LAC-DR3 catalogue. The model was trained on well-identified sources (FSRQs and BL Lacs) and subsequently applied to BCUs, providing a more complete and probabilistic characterization of the blazar population. Our main findings can be summarized as follows:

\begin{itemize}
    \item The classification probabilities assigned by \textit{TabPFN} are consistent with known blazar properties, revealing a smooth transition from FSRQs to BL Lacs, notably through an intermediate population of CLBs. This observation supports an evolutionary scenario in which blazars gradually evolve from FSRQs to HBLs due to the depletion of circumnuclear material \citep{cavaliere_2002_blazar, boettcher_2002_blazar_evolution}, as also suggested by \citet{kang_2024_clb}.
    
    \item The use of the continuous probability score $P_{\mathrm{BLL}}$ allows for a flexible characterization of sources within the intermediate range $0.2 \leq P_{\mathrm{BLL}} \leq 0.9$. This avoids rigid, pre-defined classification schemes (such as \texttt{SED\_class}) and is further strengthened by bootstrap-based uncertainty estimates that quantify the reliability of each prediction.

    \item Interestingly, CLBs consist of a heterogeneous mix of FSRQs, BL Lacs, and BCUs, most of which are LSP sources. This suggests that CLBs may retain FSRQ-like properties, such as a thin and efficient accretion disk, even during this transitional phase.

    \item By segmenting the latent space based on $P_{\mathrm{BLL}}$, we uncover three distinct populations with respect to key physical properties: gamma-ray photon index, redshift, accretion disk luminosity, and Compton dominance. In each of these features, CLBs consistently occupy an intermediate regime, supporting the hypothesis of a continuous cosmological evolution from efficient to inefficient accretion.

    \item This intermediate region is significantly populated by CLB candidates identified in recent studies \citep{kang_2024_clb, Kang2025_CLB_Mclust}, further validating the physical relevance of our latent space representation and reinforcing the robustness of our conclusions.

\end{itemize}

A central strength of our approach lies in the combination of the latent space representation and the scalar $P_{\mathrm{BLL}}$ probability. This enables an insightful visualization of the relationship between blazar type and physical properties. 

The latent representation provides a basis for inferring missing values, such as redshift or \texttt{SED\_class}, by examining a source’s position within the continuous distribution of known observational properties. This could serve as a prior for sources with incomplete data.

Ultimately, we propose a visualization-based framework that supports a possible unified evolutionary scheme for blazars, underscoring the power of combining machine learning with physical interpretation.

Nevertheless, it is important to note that uncertainties in feature measurements were not incorporated in the current implementation, as neither \textit{TabPFN} nor \textit{UMAP} natively accounts for uncertainties. Future work could implement this approach by incorporating observational errors with an updated method, by applying our current method to other blazar catalogues, and by integrating complementary data into our current approach (e.g., spectroscopic redshift, polarization, or variability measurements). These enhancements would provide even stronger constraints on the proposed evolutionary scenario from FSRQs to BL Lacs via CLBs.

Of particular interest, and to the best of our knowledge, this work represents the first application of a pre-trained foundation model for tabular data to the field of high-energy astrophysics. We showcase the strength of this approach, which requires no model-specific training or optimization, apart from minimal hyper-parameter tuning, and which yet provides fast and interpretable results.

In conclusion, this study demonstrates how modern deep learning models, such as \textit{TabPFN}, when guided by astrophysical understanding, can offer powerful new ways to explore the nature and evolution of blazars.

\begin{acknowledgements}
The authors acknowledge the support provided by Université Paris Cité to Prof. Yvonne Becherini through the Chaire Professorale IdEX, as well as support from the Data Intelligence Institute of Paris (diiP) at Université Paris Cité.
We also thank Maximilian Eff for carefully reading the final manuscript and offering valuable feedback. We thank the anonymous referee for the useful comments, which helped clarify some concepts in the paper. Minor language edits were assisted by AI tools.
\end{acknowledgements}

\bibliographystyle{aa}

\begin{appendix}

\onecolumn 

\section{Feature visualization}
\label{feat_visualization}

\begin{figure}[htbp]
    \centering
    \includegraphics[width=0.95\textwidth]{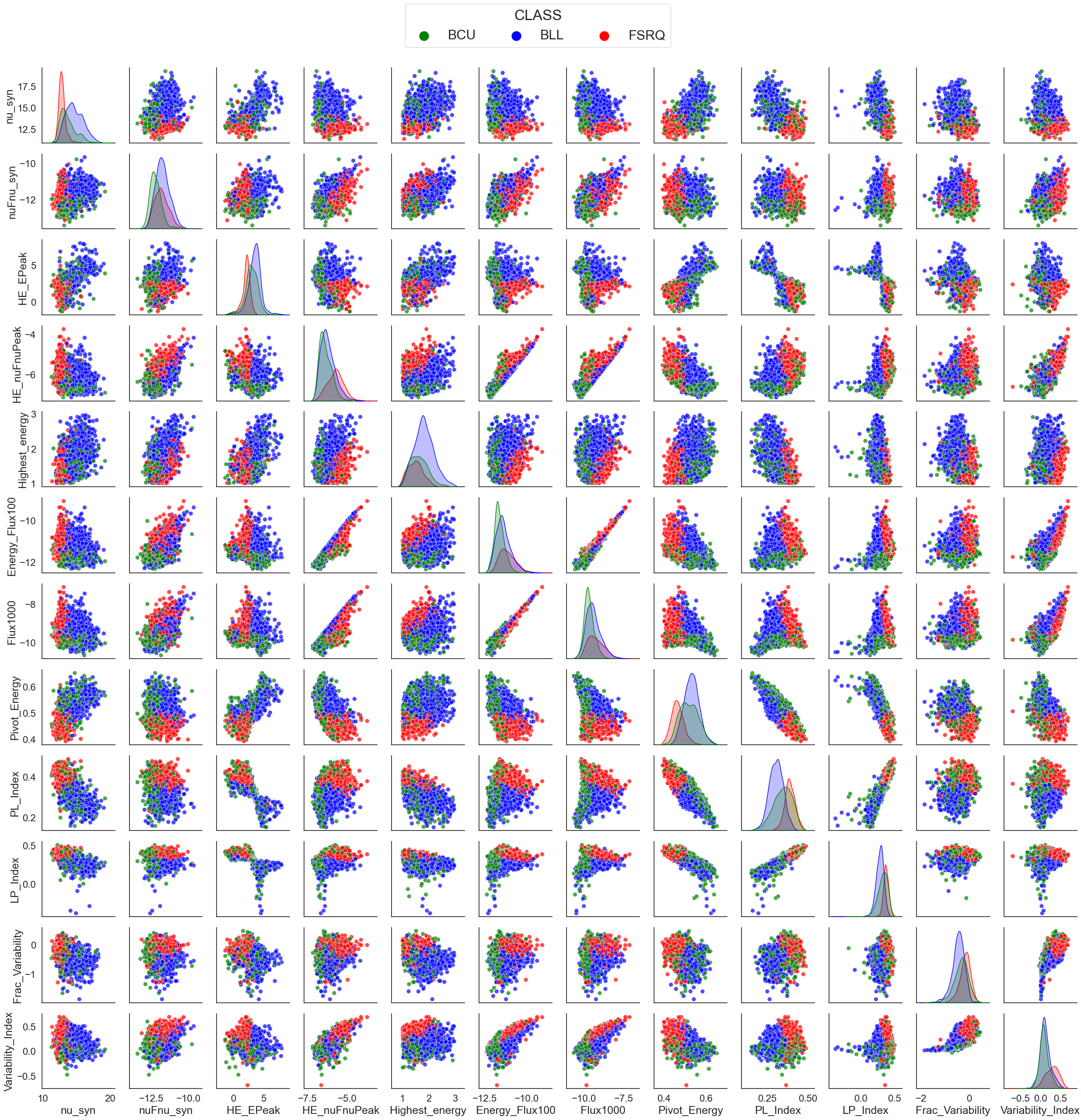}
    \caption{Visual representation of the feature space for the key 4LAC-DR3 parameters. A $\log_{10}$ transformation is applied. Missing values are not shown.}
    \label{fig:pairplot_4LAC_features}
\end{figure}

\twocolumn 

\clearpage

\section{Training strategy and optimization}\label{training_opti}
\subsection{Data split and cross-validation}\label{split_cv}

For our dataset (4LAC-DR3), we initially divide the samples into two subsets: one subset, containing only FSRQs and BL Lacs, is used for training and testing, while the other subset, consisting solely of BCUs, is reserved for subsequent predictions. Thus, the dataset of FSRQs and BL Lacs is used to train models for this classification end, based on their physical properties. Once trained, the models can then be applied to unseen data (the BCU sample) to classify uncertain blazars. A straightforward approach to training these models is to split the FSRQ and BL Lac dataset into two subsets: a training set, typically comprising 80\% of the data, used to learn patterns, and a test set, consisting of the remaining 20\%, used to evaluate the model’s performance. However, evaluating performance on a single test set might not adequately represent the entire dataset. This approach can introduce significant variability in results and produce models that fail to generalize well, increasing the risk of over-fitting.

To ensure robust training and evaluation, we employ cross-validation, a widely used technique in machine learning that involves training and testing a model multiple times on different subsets of the data. In our study, we use a method called \textit{Stratified K-Fold Cross-Validation} \citep{mahesh_2023_stratifiedkfold} to reliably evaluate our models. This involves dividing the data into $K$ equal parts. The model is trained on $K-1$ parts and then tested on the remaining part. We repeat this process $K$ times, each time testing on a different part. Finally, the overall performance is calculated by averaging the results from all $K$ tests. Unlike standard K-Fold cross-validation, Stratified K-fold Cross-Validation ensures that class proportions are maintained across all folds. This is particularly useful in our case, as in 4LAC-DR3, BL Lacs (1458) are more numerous than FSRQs (792). As we will see in the next sections, cross-validation is crucial not only for evaluating classification performance but also for the optimization of the model parameters (in a procedure called hyper-parameter tuning) and for the selection of the most relevant features for classification.

\subsection{Data preparation through feature optimization}
\label{FeatureOptimization}

Before obtaining the final classification results, it is crucial to refine the dataset by removing features that do not significantly help in classifying FSRQs and BL Lacs. Optimizing the dataset offers several advantages. Firstly, using a smaller set of relevant features helps the model to make better predictions by avoiding the possible memorization of irrelevant details (over-fitting), which improves accuracy on new data. Secondly, it makes the results easier to understand and explain, since fewer and more meaningful features provide clearer insights into the relation between the input and the output. Finally, it speeds up the analysis: fewer features mean lower memory usage and faster computations. For instance, in  \textit{XGBoost}, this results in fewer splits in decision trees, resulting in faster execution.

In this section, we present the data preparation process for the training task. This includes:
\begin{enumerate}
    \item Simple feature engineering, transforming raw data into meaningful features to optimize learning.
    \item Feature scaling, which standardizes features to ensure that they contribute equally to the model.
    \item Feature importance evaluation, by using permutation methods coupled with cross-validation to assess the significance of each feature in predicting the target variable.
\end{enumerate}

When dealing with features that span multiple orders of magnitude, like in our case, a $\log_{10}$ transformation can be an effective preprocessing step. This transformation compresses the scale of variables with wide ranges, reducing the influence of outliers while preserving the relative relationships between values. Therefore, to mitigate skewness and potentially improve the classification performance, a $\log_{10}$ transformation is applied to all features. Note that the $\log_{10}$ transformation, implemented using numpy, is applied only to numerical data, automatically ignoring missing values. During model training, missing values are explicitly converted to NaNs ("Not a Number"), allowing the classification algorithms to handle them appropriately within the learning process. After applying the logarithmic transformation, standardizing the data to have zero mean ($\mu = 0$) and unit variance ($\sigma = 1$) further normalizes the features, ensuring they contribute equally to subsequent analyses or modelling techniques. This step ensures well-defined distributions closer to normality, which is essential for foundation models like  \textit{TabPFN}, as they have been pre-trained on normalized distributions \citep[see][Section C.2.1]{hollmann_2023_tabpfn}. However, tree-based models like \textit{XGBoost} do not require such feature scaling approaches, as they inherently handle different feature scales, although applying them can sometimes improve convergence and stability. Therefore, we apply both transformations (logarithmic + standardization) for  \textit{TabPFN}, while for \textit{XGBoost}, we only apply the $\log_{10}$ transformation.

To further evaluate the significance of our features, we apply a permutation importance method combined with cross-validation. The principle behind this method is to evaluate how much the model's performance is affected when the values of a given feature are randomly shuffled. This process helps determine the contribution of each feature for the classification task: if the model performance drops, the feature is important, while if the performance remains unchanged, the feature is less relevant or redundant. To ensure stability, the permutation process is repeated multiple times averaging the results to obtain a robust estimate of feature importance. For  \textit{XGBoost}, this feature importance analysis is performed in parallel with stratified cross-validation across the full dataset. However, as mentioned in Section \ref{classif_models}, \textit{TabPFN} is designed to operate in a zero-shot manner, meaning it does not require multiple training iterations. 
Consequently, applying stratified cross-validation for feature optimization on \textit{TabPFN} would result in high computational costs and long execution times, even with parallelization. To mitigate this, feature importance analysis for \textit{TabPFN} was performed without cross-validation. This approach provides a sufficiently accurate estimation of the most relevant features for distinguishing FSRQs from BL Lacs, while maintaining computational efficiency.

Feature optimization results for 4LAC-DR3, with \textit{TabPFN} (top) and \textit{XGBoost} (bottom), can be seen in Fig.~\ref{fig:4lac_feat_importance}. Since we have an imbalanced dataset (FSRQs vs BL Lacs), the accuracy (percentage of correctly classified sources) is not the most reliable metric to evaluate classification performance. We therefore consider the following key metrics \citep[see][]{sklearn_model_eval} that better reflect the model's ability to distinguish between both source types:

\begin{itemize}
    \item Balanced accuracy which takes into account the accuracy for each class independently, providing a better evaluation in case of class imbalance.
    \item ROC-AUC (receiver operating characteristic - area under the curve) measuring how well the model separates FSRQs from BL Lacs. It is based on the ROC curve, which plots the true positive rate (TPR) against the false positive rate (FPR). A ROC-AUC score of 0.5 means the model is making random guesses, while a score of 1 indicates a perfect classification.
    \item {$\mathrm{F_{1}}$-score} that balances precision (how many of the sources classified as BL Lacs are truly BL Lacs) and recall (how many of the actual BL Lacs are correctly identified). It is particularly useful when dealing with imbalanced datasets, as it avoids misleadingly high performance due to majority class dominance.
\end{itemize}

\subsection{Identifying key features in the 4LAC catalogue}\label{identify_feat}

Based on these metrics, we observe that the most relevant features for distinguishing FSRQs from BL Lacs in the 4LAC catalogue includes :

\begin{itemize}
     \item The spectral indices that represent the high-energy SED spectral behaviour, captured by the power-law and log-parabolic indices (\texttt{PL\_Index}, \texttt{LP\_Index}). These features separate the two classes by encoding distinctions in their gamma-ray spectral shapes.
     \item The synchrotron peak which represents the synchrotron peak frequency (\texttt{nu\_syn}) and reflects the difference in the jet emission physics.
     \item The pivot energy which represents the reference energy (\texttt{Pivot\_energy}) 
     at which the gamma-ray spectral parameters are most tightly constrained. This feature serves as a reference energy, reflecting differences in the inverse Compton peak energies of FSRQs and BL Lacs. FSRQs typically exhibit lower pivot energies, while BL Lacs often show higher pivot energies.
    \item The source variability which represents the flux variability of the source (quantified by \texttt{Variability\_Index}) and often offers supplementary discriminative power.
\end{itemize}

It is worth noting that both models account for potential correlations between features. In the case of \textit{TabPFN}, this is handled during the pre-training phase through a technique called \textit{block-wise feature sampling}, where adjacent features are grouped to reflect the correlated behaviour observed in the data. This approach allows for the inclusion of a wider range of features during the feature selection process without imposing restrictions on their correlation.

\begin{figure*}[ht]
    \centering
    \includegraphics[width=1\textwidth]{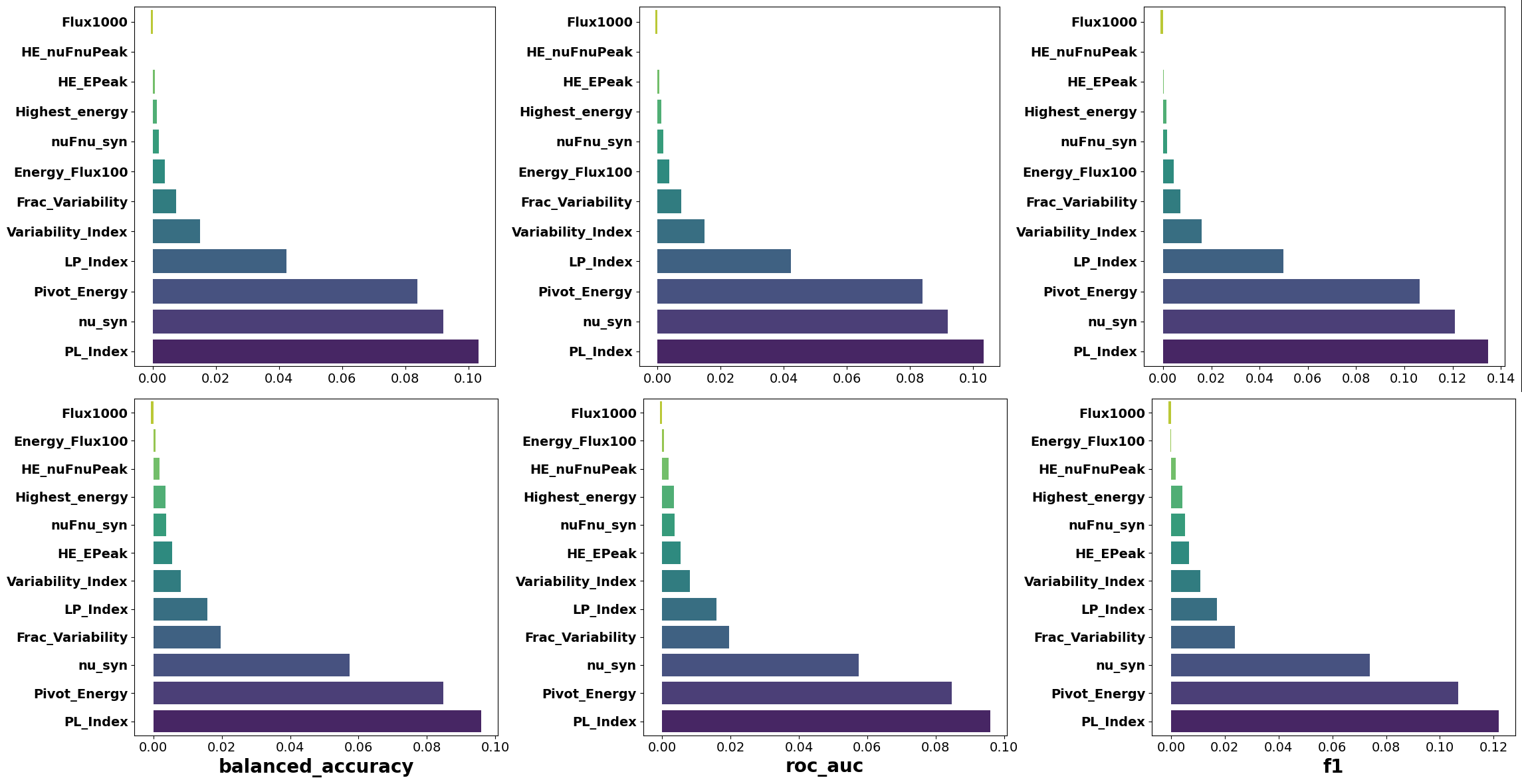}
    \caption{Feature importance results for different classification metrics (balanced\_accuracy, roc\_auc and f1) with \textit{TabPFN} (top panel) and \textit{XGBoost} (bottom panel).}
    \label{fig:4lac_feat_importance}
\end{figure*}

As shown in Fig.~\ref{fig:4lac_feat_importance}, it appears that \textit{TabPFN} assigns notable importance to a broader range of features than \textit{XGBoost}.
This is likely due to the fact that \textit{TabPFN} is a more complex model, capable of capturing intricate relationships between features. Interestingly, while \texttt{PL\_Index} consistently ranks among the most important features for distinguishing FSRQs from BL Lacs in both models, information related to the energy/flux of the inverse Compton peak appears to be less relevant. In contrast, features related to variability, the synchrotron peak, and the high-energy slope seem to play a more crucial role in classification.

When using only the top key features, we observe an improvement across all evaluated metrics (balanced accuracy, ROC AUC, and $F_1$ score). Consequently, we chose to retain the 8 most important features for each model, as shown in Fig.~\ref{fig:4lac_feat_importance}. This decision is further supported by the fact that complex models like \textit{TabPFN} tend to perform better when provided with a compact yet informative set of features, as they are capable of capturing intricate patterns in the feature space.

\subsection{Fine-tuning model parameters}\label{hyper-param}
Once the data have been properly prepared, it is often crucial to fine-tune the parameters of the classification model, also known as hyper-parameters, in a process referred to as hyper-parameter optimization.

In the case of  \textit{XGBoost}, hyper-parameter tuning is essential. Since \textit{XGBoost} is based on decision trees, it has numerous hyper-parameters that influence the performance and require careful optimization to achieve the best results. 
For example, the depth of the trees controls the complexity of the model, and adjusting this value helps in avoiding over-fitting (where an overly complex model memorizes the data instead of generalizing) 
or under-fitting (where the model is too simple to identify meaningful patterns).
Similarly, key hyper-parameters such as the learning rate and the number of estimators regulate the learning process and its efficiency. Additionally, given the imbalanced nature of the dataset, adjusting the class weights during training helps to account for the fact that certain source types (e.g, BL Lacs) are overrepresented compared to others (e.g, FSRQs) in datasets like 4LAC-DR3.

On the other hand, \textit{TabPFN} is fundamentally different from  \textit{XGBoost}, as it is based on pre-trained neural networks that make instantaneous predictions without requiring specific training on the blazar catalogues. As a result, hyper-parameter optimization is unnecessary, since \textit{TabPFN} has already been optimized beforehand to adapt to a wide range of tabular datasets, considering different possible data distributions.

Consequently, we focus our hyper-parameter optimization efforts on  \textit{XGBoost}, using the Python library Optuna \citep{akiba_2019_optuna}. This tool explores the hyper-parameter space by testing different configurations, based on advanced Bayesian optimization methods. 
In parallel, we also perform cross-validation with 10 folds to ensure robust results. For each catalogue of blazars, we perform 100 trials with Optuna, leading to a selection of 10 hyper-parameters for tuning (including the number of estimators, max-depth, learning rate, and class weight), along with their optimal values. This optimization improves accuracy, as well as the roc\_auc and $F_1$ score metrics.
One advantage of this method is that it automatically adjusts class weights, addressing the imbalance in the dataset.

As a side note, we can still apply hyper-parameter optimization in the case of \textit{TabPFN} for 5 different parameters, including the \texttt{n\_estimators} (the number of estimators in the \textit{TabPFN} ensemble), \texttt{softmax$\_$temperature} (which controls the confidence of the model's predictions), and \texttt{balance\_probabilities} (which adjusts the output probabilities to treat the classes as if they were equally likely).

Nevertheless, only slight performance differences were observed across these configurations, reflecting \textit{TabPFN}’s overall robustness.

\section{Testing the quality of the model}
\label{results_test}
\subsection{Performance on FSRQs vs BL Lacs}\label{fsrq_vs_bll}

After selecting the most relevant features and optimizing the hyper-parameters to enhance the classification of FSRQs and BL Lacs, we proceed to fit the models (\textit{XGBoost} and \textit{TabPFN}) on the training data. We then check how well the model performs by testing it on different parts of the data, using  10-fold cross-validation. The performance of \textit{XGBoost} and \textit{TabPFN} are summarized in Table~\ref{tab:perf_4lacdr3}. Both models perform exceptionally well, with \textit{TabPFN} showing slightly better performance across the board. Key metrics like accuracy, precision, recall, and $F_1$-score are consistent with those reported in \citet{agarwal_2023_classif_blazars} and \citet{bhatta_2024_classif_blazars}, which primarily relied on more traditional machine learning approaches. In contrast, our use of a foundation model (\textit{TabPFN}) yields comparable performance while benefiting from its inherent ability to generalize and adapt across datasets without extensive hyper-parameter tuning.

\renewcommand{\thetable}{2}
\begin{table*}[t]
    \centering
    \caption{Classification performance of \textit{XGBoost} and \textit{TabPFN} for the sample of FSRQs and BL Lacs in the 4LAC-DR3.}
    \resizebox{\textwidth}{!}{
    \begin{tabular}{c c ccc ccc c}
        \hline
        \hline
        \textbf{Model} & \textbf{Accuracy} & & \textbf{BL Lac} & & & \textbf{FSRQ} & & \textbf{AUC} \\
        & & \textbf{Precision} & \textbf{Recall} & \textbf{$F_1$ score} & \textbf{Precision} & \textbf{Recall} & \textbf{$F_1$ score} & \\
        \hline
        \textit{XGBoost} & 0.908 $\pm$ 0.017  & 0.932 $\pm$ 0.021 & 0.931 $\pm$ 0.021 & 0.931 $\pm$ 0.013 & 0.864 $\pm$ 0.036 & 0.864 $\pm$ 0.044 & 0.863 $\pm$ 0.027 & 0.963 $\pm$ 0.011 \\
        \textit{TabPFN} & 0.916 $\pm$ 0.019 & 0.938 $\pm$ 0.021 & 0.936 $\pm$ 0.018 & 0.937 $\pm$ 0.014 & 0.874 $\pm$ 0.032 & 0.875 $\pm$ 0.045 & 0.874 $\pm$ 0.030 & 0.969 $\pm$ 0.009 \\
        \hline
    \end{tabular}
    }
    \label{tab:perf_4lacdr3}
\end{table*}

\begin{figure*}[ht]
    \centering
    \includegraphics[width=0.5\textwidth]{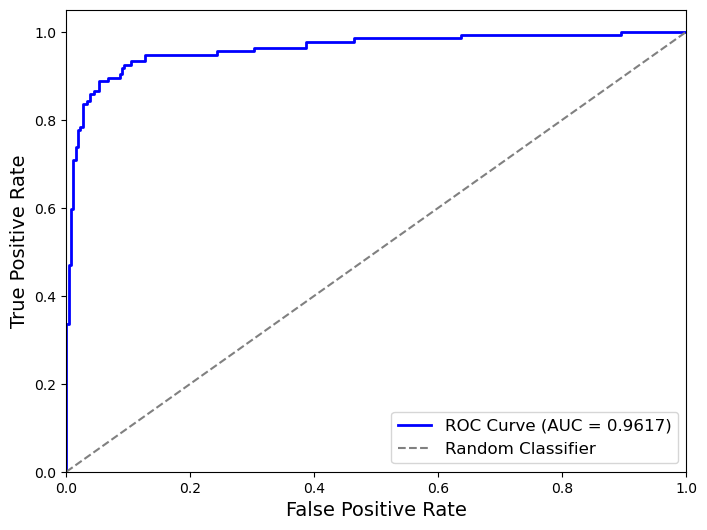}
    \caption{The ROC curve obtained with the application of \textit{TabPFN} to the sample of FSRQs vs BL Lacs in the 4LAC-DR3.}
    \label{fig:ROC_curve}
\end{figure*}

Both models achieved AUC values exceeding 0.95, with \textit{TabPFN} reaching 0.969. This result outperforms the AUC reported in recent works such as \citet{agarwal_2023_classif_blazars} and \citet{bhatta_2024_classif_blazars}, underlining the effectiveness of foundation models in tabular classification tasks involving complex astrophysical data. As mentioned previously, \textit{XGBoost} was primarily used as a strong baseline to benchmark our foundation model. The results confirm that \textit{TabPFN} consistently outperforms \textit{XGBoost}, achieving higher accuracy and AUC scores.

\begin{figure}[t]
    \centering
    \includegraphics[width=0.5\textwidth]{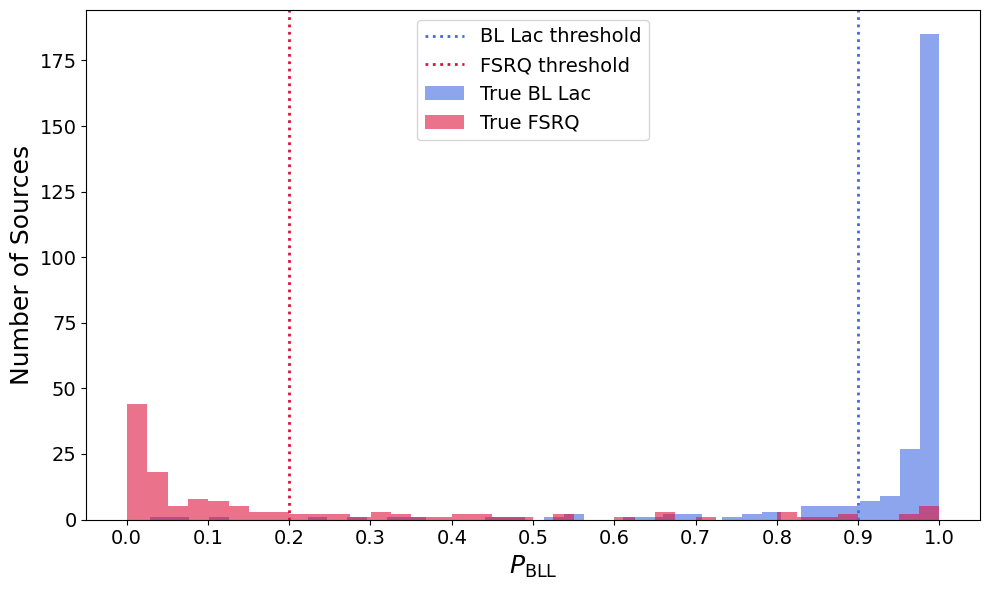}
    \caption{Distribution of BL Lac classification probabilities (\( P_{\mathrm{BLL}}\)) for all sources in the test set, as predicted by \textit{TabPFN}. The X-axis represents \( P_{\mathrm{BLL}} \), indicating the predicted probability of a source being a BL Lac, while the colour of each bar corresponds to its true classification in the catalogue. Dashed vertical lines indicate the thresholds derived from the KDE analysis, marking regions where sources are most confidently identified as either BL Lac-like (\(P_{\mathrm{BLL}} > 0.9\)) or FSRQ-like (\(P_{\mathrm{BLL}} < 0.2\)).}
    \label{fig:distri_BLL_proba}
\end{figure}

To further assess \textit{TabPFN}’s performance, we display Fig.~\ref{fig:ROC_curve} the ROC curve which illustrates the model’s ability to distinguish between FSRQs and BL Lacs. The AUC score quantifies this performance: a value close to 1 indicates that the model is highly effective in separating the two classes. Additionally, Fig.~\ref{fig:distri_BLL_proba} shows the distribution of BL Lac classification probabilities ($P_{\mathrm{BLL}}$) as predicted by \textit{TabPFN}. These probabilities were obtained for all sources in a test sample, consisting exclusively of BL Lacs and FSRQs.

In these plots, each bar represents several sources, with the colour corresponding to the true classification from the 4LAC catalogue. The X-axis indicates the predicted probability $P_{\mathrm{BLL}}$, which quantifies the likelihood of a source being classified as a BL Lac. A well-separated distribution with peaks near 0 and 1 suggests that the model confidently distinguishes between FSRQs and BL Lacs. High-density peaks are observed at both ends of the distribution, around \( P_{\mathrm{BLL}} = 0 \) (for FSRQs) and \( P_{\mathrm{BLL}} = 1 \) (for BL Lacs). Sources with $P_{\mathrm{BLL}}$ near 0 correspond to the true FSRQs (as indicated by the red colour). Similarly, sources with \( P_{\mathrm{BLL}} \) close to 1 correspond to the true BL Lacs (as indicated by the blue colour). Only a small fraction of true BL Lacs is assigned a low \( P_{\mathrm{BLL}} \), and conversely, only a small fraction of true FSRQs has a high \( P_{\mathrm{BLL}} \). This result highlights \textit{TabPFN}'s ability to correctly distinguish between the two classes. Notably, the performance reported in Table~\ref{tab:perf_4lacdr3}, which is lower for the FSRQ class than for BL Lacs, is due to a higher proportion of misclassified FSRQs, as illustrated in Fig.~\ref{fig:distri_BLL_proba}. These correspond to sources originally classified as FSRQs in the 4LAC catalogue but exhibiting observational properties identical to BL Lacs (i.e., a high $P_{\mathrm{BLL}}$ value). Finally, we apply a KDE to the distribution of \(P_{\mathrm{BLL}}\) values in order to locate the probability density peaks associated with each population (\texttt{kde\_peak}). This allows us to define two classification thresholds: one above which sources are considered BL Lac-like, and another below which sources are considered FSRQ-like. To ensure that the thresholds remain close to the respective high-density peaks and avoid ambiguous cases near \(P_{\mathrm{BLL}} \sim 0.5\), we adopt a conservative criterion based on a small shift from the peak, defined as \(\texttt{kde\_peak} \pm 0.5 \times \texttt{std}\), where \texttt{std} is the standard deviation of the distribution around the peak. The resulting thresholds are shown in Fig.~\ref{fig:distri_BLL_proba}, with a BL Lac-like threshold set at \(P_{\mathrm{BLL}} = 0.91\) (blue dashed line) and a FSRQ-like threshold at \(P_{\mathrm{BLL}} = 0.19\) (red dashed line) which we rounded to 0.9 and 0.2.

\begin{figure}[t]
    \centering
    \includegraphics[width=0.5\textwidth]{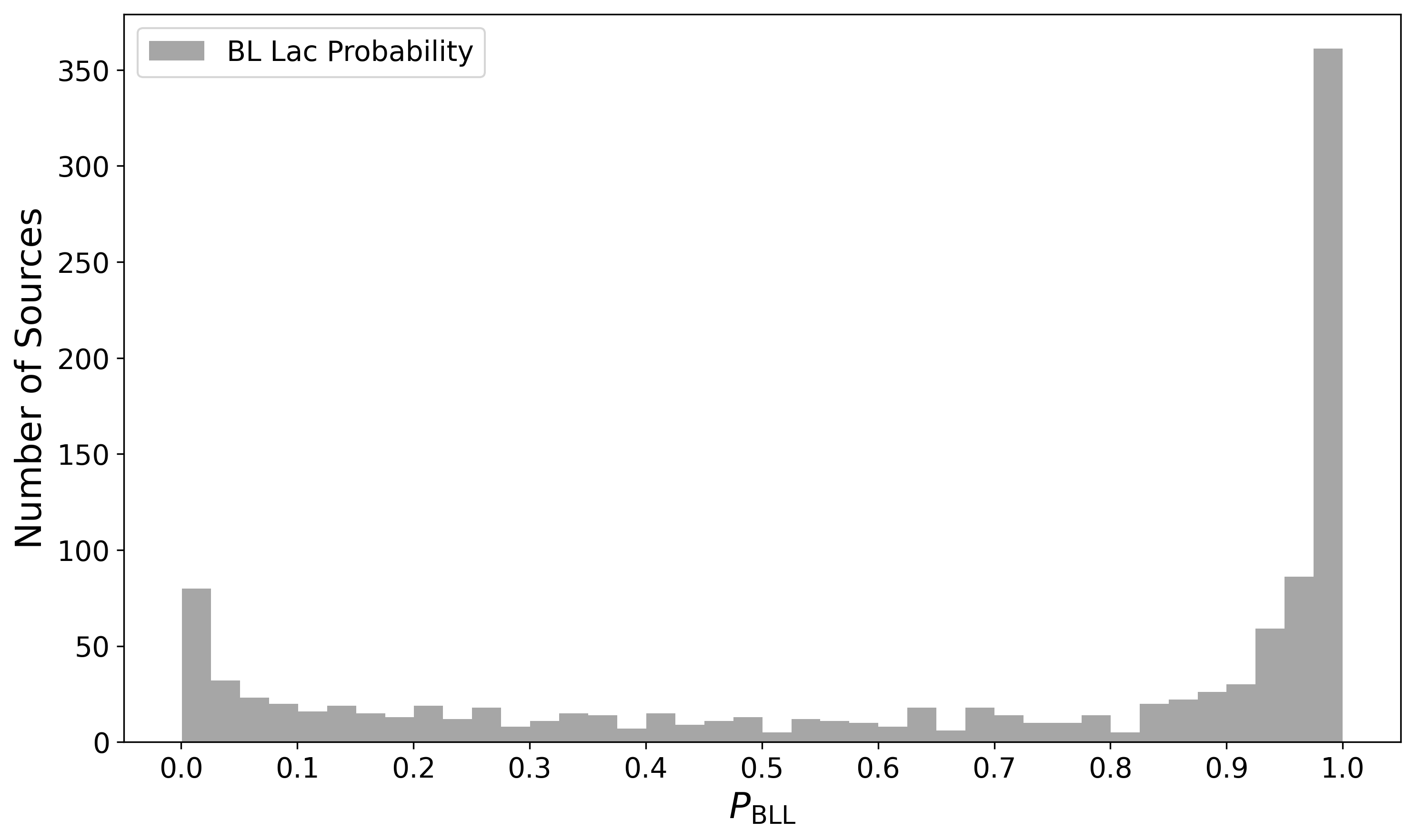}
    \caption{Distribution of BL Lac classification probabilities (\( P_{\mathrm{BLL}} \)) for all BCUs, as predicted by \textit{TabPFN}. The predictions were obtained using a bootstrap method with 50 instances of \textit{TabPFN} trained on different subsets of the data.}
    \label{fig:distri_BCU_proba}
\end{figure}

\subsection{Application on BCUs}\label{application_bcu}

After optimizing \textit{TabPFN} on the FSRQ/BL Lac subsets of the 4LAC-DR3 catalogue, we apply the model to classify BCUs. Fig.~\ref{fig:distri_BCU_proba} shows the distributions of the predicted BL Lac probability, \(P_{\mathrm{BLL}}\), for these BCUs. As in Section~\ref{fsrq_vs_bll}, many predictions cluster near 0 or 1, suggesting that the model can efficiently classify a significant proportion of BCUs as either FSRQs or BL Lacs. 

To quantify the reliability of these predictions, we estimate uncertainties via an ensemble approach. Specifically, we generate 50 bootstrapped training sets (resampling data with replacement) and train a separate \textit{TabPFN} instance, each with a unique random initialization, on each data sample. For every BCU, we then average its 50 probability estimates (\(\langle P_{\mathrm{BLL}} \rangle\)) and compute their standard deviation as an uncertainty measure. The probability distribution in Fig.~\ref{fig:distri_BCU_proba} was obtained using this method, aggregating predictions from 50 \textit{TabPFN} models trained on different data subsets. This approach helps us estimating how much predictions can vary without relying on strict assumptions about the error distribution. Therefore by simulating different training scenarios, we can estimate the uncertainty in the predictions. This approach is particularly valuable for methods like \textit{TabPFN}, which does not provide explicit Bayesian-style uncertainty estimates out of the box.

\begin{figure}[t]
    \centering
    \includegraphics[width=0.5\textwidth]{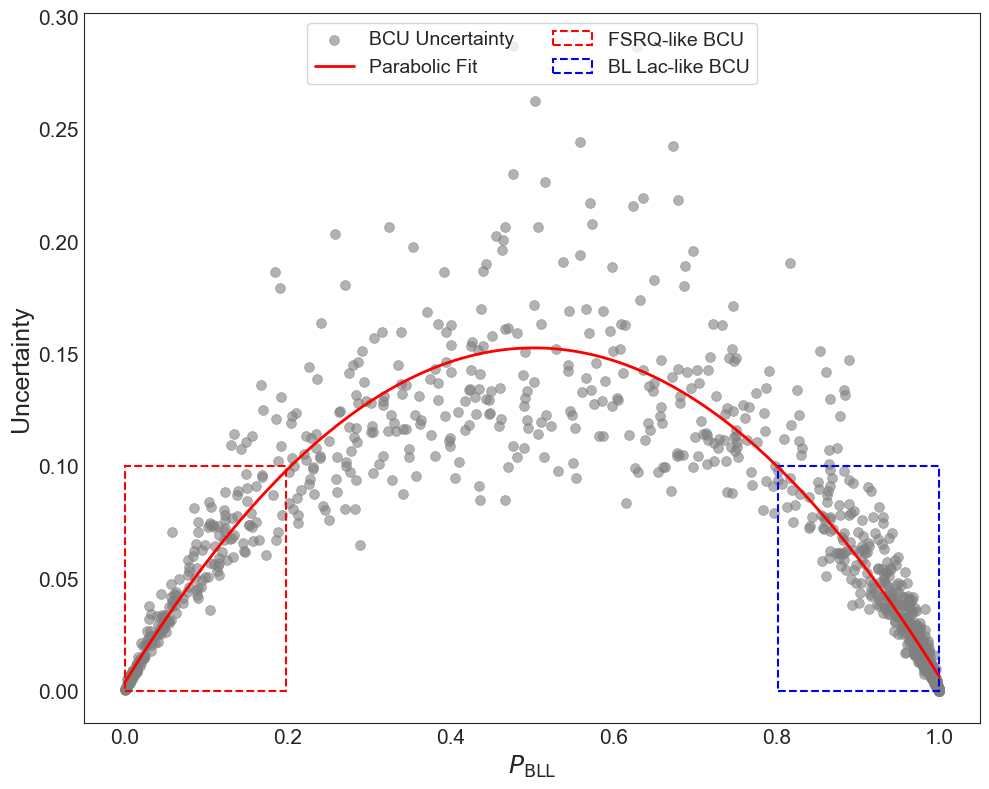}
    \caption{Average bootstrap BL Lac probabilities versus associated uncertainties for 4LAC-DR3 BCUs. A parabolic fit is overlaid to emphasize the trend and highlight potential outliers. The red and blue boxes enclose BCU sources with a high probability of being FSRQ and BL Lac, respectively, both exhibiting low uncertainty.}
    \label{fig:uncertainty_vs_proba}
\end{figure}

To further analyse these uncertainties, we plot them as a function of the averaged predicted probabilities (\( \langle P_{\mathrm{BLL}} \rangle \)) in Fig.~\ref{fig:uncertainty_vs_proba}. A parabolic fit is overlaid to highlight the general trend. As expected, the highest uncertainties occur when the predicted probability is close to 0.5. This is because sources with a probability near 50\% of being a BL Lac (\( P_{\mathrm{BLL}} \approx 0.5 \)) also have an equal probability (1 - 0.5 = 50\%) of being an FSRQ. Such cases indicate that, based on their physical properties, the model struggles to confidently classify them as either type. In particular, we observe that a significant number of BCUs exhibit uncertainties greater than 25\% when their predicted BL Lac probability is close to 50\%. These outliers may represent a transitional class between FSRQs and BL Lacs, exhibiting properties that slightly deviate from both categories. To further highlight the classification reliability, a red box has been drawn around BCUs with a low probability of being BL Lac ($P_{\mathrm{BLL}}$ < 0.2), meaning they are most likely FSRQs. Conversely, a blue box encloses BCUs with a high probability of being BL Lacs ($P_{\mathrm{BLL}}$ > 0.8). These boxes also account for uncertainties by including only sources with prediction uncertainties below 0.1. Therefore, to study BCUs that closely resemble either FSRQs or BL Lacs, we focus on the sources within these respective boxes, while avoiding those located outside of them, as they exhibit greater classification ambiguity.

\FloatBarrier

\end{appendix}
 
\end{document}